\documentclass[aps,pra,preprint,floatfix,longbibliography, superscriptaddress]{revtex4-1}

\usepackage{natbib}
\usepackage{xcolor}
\usepackage{appendix}
\newcommand*{\colorboxed}{}
\def\colorboxed#1#{%
	\colorboxedAux{#1}%
}
\newcommand*{\colorboxedAux}[3]{%
	\begingroup
	\colorlet{cb@saved}{.}%
	\color#1{#2}%
	\boxed{%
		\color{cb@saved}%
		#3%
	}%
	\endgroup
}

\usepackage{pifont}

\usepackage{float}
\usepackage{epsfig}
\usepackage{bm}
\usepackage[matrix,frame,arrow]{xy}
\usepackage[applemac]{inputenc}
\usepackage[T1]{fontenc}
\usepackage{lmodern}
\usepackage[english]{babel}
\usepackage{amsmath}
\usepackage{ae}
\usepackage{amssymb}
\usepackage{color}
\usepackage{graphicx}
\usepackage{bbm}
\usepackage{url}
\usepackage{nomencl}
\usepackage{subfigure}
\usepackage{slashed}

\usepackage{booktabs}

\newcommand{\figref}[1]{\mbox{Fig.~\ref{#1}}}

\renewcommand{\eqref}[1]{\mbox{Eq.~(\ref{#1})}}

\newcommand{\be}{\begin{equation}}
\newcommand{\ee}{\end{equation}}
\newcommand{\bea}{\begin{eqnarray}}
\newcommand{\eea}{\end{eqnarray}}

\usepackage{xr}
\usepackage[colorlinks]{hyperref}
\hypersetup{%
    plainpages=true,
    breaklinks=true,
    hypertexnames=false,
    pageanchor=true,
    colorlinks=true,
    linkcolor={blue},
    citecolor={red},
    urlcolor={blue},
    anchorcolor={black}
}

\makeatletter

\newcommand{\checknextarg}{\@ifnextchar\bgroup{\gobblenextarg}{and}}
\newcommand{\gobblenextarg}[1]{,\! (\ref{#1})\@ifnextchar\bgroup{\gobblenextarg}{]}}

\begin{document}
	
	
\title{Gauge invariance of the Dicke and Hopfield models}
%
%
%
%
	\author{Luigi Garziano}
	\affiliation{Theoretical Quantum Physics Laboratory, RIKEN Cluster for Pioneering Research, Wako-shi, Saitama 351-0198, Japan}
%
\author{Alessio Settineri}
\affiliation{Dipartimento di Scienze Matematiche e Informatiche, Scienze Fisiche e  Scienze della Terra,
Universit\`{a} di Messina, I-98166 Messina, Italy}

\author{Omar Di Stefano}
\affiliation{Theoretical Quantum Physics Laboratory, RIKEN Cluster for Pioneering Research, Wako-shi, Saitama 351-0198, Japan}
\affiliation{Dipartimento di Scienze Matematiche e Informatiche, Scienze Fisiche e  Scienze della Terra,
	Universit\`{a} di Messina, I-98166 Messina, Italy}

\author{Salvatore Savasta}
\affiliation{Dipartimento di Scienze Matematiche e Informatiche, Scienze Fisiche e  Scienze della Terra,
	Universit\`{a} di Messina, I-98166 Messina, Italy}
\email[corresponding author: ]{ssavasta@unime.it}

\author{Franco Nori}
	\affiliation{Theoretical Quantum Physics Laboratory, RIKEN Cluster for Pioneering Research, Wako-shi, Saitama 351-0198, Japan} \affiliation{Physics Department, The University
		of Michigan, Ann Arbor, Michigan 48109-1040, USA}

	\begin{abstract}

		The Dicke  model, which describes the
		dipolar coupling between $N$ two-level atoms and a quantized electromagnetic field, seemingly violates gauge invariance in the presence of ultrastrong light–matter coupling, a regime that is now experimentally accessible in many physical systems.
Specifically, it has been shown that, while the two-level approximation can work well in the dipole gauge, the  Coulomb gauge fails to provide the correct spectra in the ultrastrong coupling regime.
Here we show that, taking into account  the nonlocality of the  atomic potential induced by the two-level approximation, gauge invariance is fully restored for arbitrary interaction strengths, even in the $N \to \infty$ limit. 
Finally, we  express  the Hopfield model, a general description based on the quantization of a linear dielectric medium, in a manifestly gauge invariant form, and show that the Dicke model in the dilute regime can be regarded as a particular case of the more general Hopfield model.

	\end{abstract}
	

	
	
\maketitle	


\section{Introduction}

Models describing the interaction between one or few modes of the electromagnetic field in a resonator and individual or ensembles of few levels  atoms are a cornerstone of quantum optics. The simplest examples are the quantum Rabi \cite{Rabi1936,Rossatto2017, DiStefano2017} and the Dicke Hamiltonians \cite{Dicke1954,Brandes2005,Garraway2011,Kirton2019} describing, respectively,  the interaction of  a single-mode bosonic field with a two-level atom, and with an ensemble of $N$ two-level atoms. Their simplified version obtained after the rotating wave approximation are the Jaynes-Cummings and Tavis-Cummings models \cite{Fink2009, Feng2015}, respectively. 

\begin{figure}[H]  
	\centering
	\includegraphics[scale=0.40]{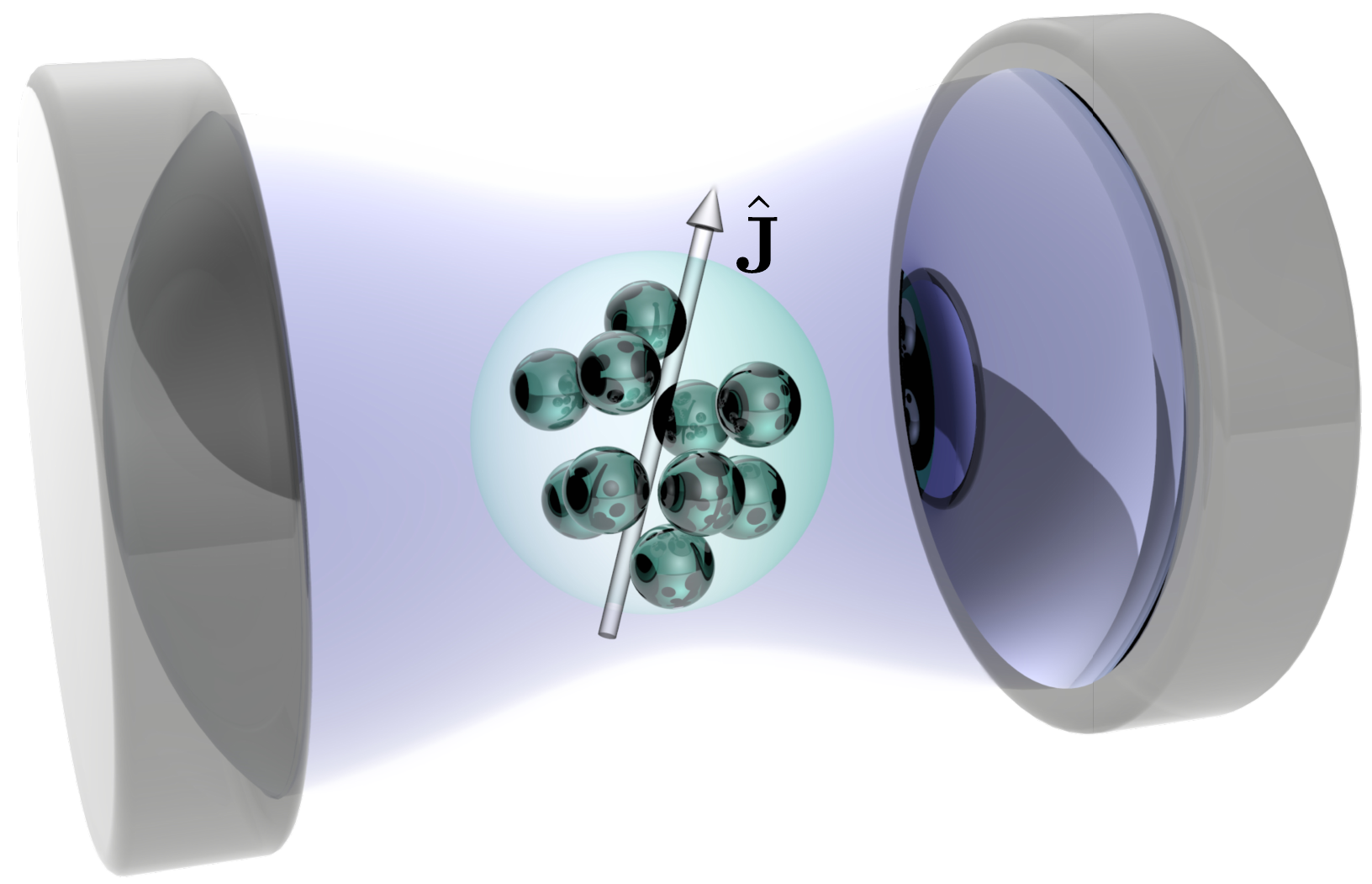}  
	\caption{Sketch of an optical resonator
		coupled to $N$ identical, distinguishable, quantum emitters.  We consider two-level emitters, that can be described by means of collective operators $\hat J_\alpha$  with $\alpha \equiv \{ x,y,z\}$,  which obey the angular momentum commutation relations (with cooperation number $j= N/2$). These atoms interact with a bosonic mode of frequency $\omega_c$ via a dipole interaction.	The resulting normalized collective coupling strength scales $\propto  \sqrt{N}$.} 
	\label{fig1}
\end{figure}

Recently, it has been argued that truncations of the atomic Hilbert space, to obtain a two-level description, violate the gauge principle \cite{DeBern2018, DeBernardis2018, Stokes2019}. Such violations become particularly relevant  in the case of ultrastrong (USC) light–matter coupling, a regime, now experimentally accessible in many physical systems, in which the coupling strength is comparable to the transition energies in the system \cite{Kockum2018,Forn-Diaz2018}.
In particular, it has been shown that, while in the electric dipole gauge the two-level approximation can be performed as long as the Rabi frequency remains much smaller than the energies of all higher-lying levels, it can drastically fail in the Coulomb gauge, even for systems with an extremely anharmonic spectrum \cite{DeBernardis2018}. 
The Dicke Hamiltonian, a model of key importance  for the description of collective effects in quantum optics, shares analogous worrying problems, not only in the presence of a small number $N$ of atoms, but also in the so-called dilute regime, where $N \to \infty$, while the coupling strength between the field and the resulting collective excitations remains finite \cite{DeBernardis2018}. Examples of realizations of the Dicke model in the  USC dilute regime include intersubband  organic molecules \cite{Schwartz2011,Kena-Cohen2013,Gubbin2014,Mazzeo2014,Gambino2014,Genco2018}, intersubband polaritons \cite{Anappara2009, Gunter2009, Todorov2010,Askenazi2014}, and Landau polaritons  \cite{Scalari2012,Maissen2014, Zhang2016a,Bayer2017,Li2018}.

In quantum electrodynamics, the choice of gauge influences the form of light–matter interactions. However, gauge invariance implies that all physical results should be independent of this formal choice. As a consequence, the observation that the quantum Rabi and Dicke model provide gauge-dependent energy spectra casts doubts on the reliability of these widespread descriptions.

The source of these gauge violations has been recently identified and  a general method for the derivation of light–matter Hamiltonians
in truncated Hilbert spaces, able to  produce gauge-invariant physical results, even for extreme light–matter interaction regimes has been proposed \cite{DiStefano2019}.
According to the gauge principle, the coupling of the matter system with the electromagnetic field is introduced by  the minimal replacement rule $\hat {\bf p} \to \hat {\bf p} - q \hat {\bf A}$, where $\hat {\bf p}$ is the momentum of an effective particle, $\hat{\bf A}$ is the vector potential of the field, and $q$ is the charge.
It has benn known for decades that approximations in the description of a quantum system, with space truncation can give rise to \textit{nonlocal} potentials which can always be expressed as potentials depending on \textit{both} position and momenta: $V({\bf r}, \hat {\bf p})$ \cite{Starace1971}.  In these cases,  in order not to ruin the gauge principle, the minimal coupling replacement has to be applied not only to the kinetic energy of the particles in the system, but also to the nonlocal potentials in the effective Hamiltonian of the matter system \cite{Starace1971, Girlanda1981,Ismail-Beigi2001}. Once this procedure is applied, it is possible to obtain gauge-invariant models, even in the presence of extreme light-matter interaction regimes \cite{DiStefano2019, Settineri2020}.
This method has been applied to obtain a quantum Rabi model satisfying the gauge principle \cite{DiStefano2019}. In the following, we will refer to models not violating gauge invariance  as gauge-invariant (GI) models, even if the form of the Hamiltonians change after a gauge transformation.  The generalization to $N$ two-level systems (Dicke model) is briefly discussed in the supplementary material of Ref.~\cite{Settineri2020}. The resulting GI quantum Rabi and Dicke Hamiltonians in the Coulomb gauge differ significantly in form from the standard ones and both contain field operators to \textit{all} orders.

Here, after revisiting the derivation of the GI Dicke model,  we derive the corresponding dilute regime, also know as thermodynamic limit \cite{Emary2003,Lambert2004,Nathan2017,Nathan2018}.
In such a limit, applying the Holstein-Primakoff transformation \cite{Holstein1940}, the standard Dicke Hamiltonians in the dipole and in the Coulomb gauges, both bilinear in the bosonic operators, are obtained (see, e.g., \cite{Emary2003}). Such Hamiltonian can be diagonalized exactly, using a multimode Bogoliubov transformation. However, it has been shown that the effective Hamiltonians in the Coulomb and dipole gauge give rise to polariton eigenfrequencies  (modes) which can significantly differ for large coupling strengths \cite{DeBernardis2018}.
Although the form of the gauge-invariant Dicke model contains field operators to all orders and appears very different from a bilinear Hamiltonian, we show that, in the thermodynamic limit, a bilinear Hamiltonian very similar to the standard one is obtained. Specifically, the resulting Dicke Hamiltonian in the Coulomb gauge only differs from the standard one for the coefficient of the diamagnetic term (proportional to $\hat {\bf A}^2$). However, we show that such a difference is sufficient to restore gauge invariance.

Another widespread description of the interaction between the quantized electromagnetic field  and collective excitations is the Hopfield model \cite{Hopfield1958}. This model was initially introduced to describe the interaction of the electromagnetic field with an harmonic resonant polarization density of a 3D dielectric crystal. Nowadays, it is used to describe the interaction between free or confined light and different kinds of collective excitations, as optical phonons, excitons in nanostructures, magnons, plasmonic crystals, which can be described as bosonic fields.
We compare the (GI) Dicke and the Hopfield models and apply to the latter the concepts derived for obtaining the first. In  doing so, we provide a method to derive in a simple way manifestly gauge-invariant  Hopfield models, having only knowledge about the matter polarization field.

\section{The Dicke Model with finite number of dipoles}

For the following analysis, we consider a generic setting
as shown in \figref{fig1}, where a finite number of electric dipoles are coupled
to the single mode of the electromagnetic field in a resonator (see, e.g., \cite{DeBernardis2018}).
The dipoles can be modeled as effective particles of mass $m$ in potentials $V(x_i)$, where $x_i$ is the separation between the charges $q$ and $-q$ of the $i$-th dipole. In the absence of any dipole-dipole interaction, and of the interaction with the electromagnetic field, the Hamiltonian describing a system of $N$ effective particles can be written as $\hat H^{(N)}_0 = \sum_{i=1}^{N} \hat H^{(i)}_{0}$,
where
\be
	\hat H^{(i)}_{0} = \frac{ \hat p_i^2}{2 m} + V(x_i)\, .
\ee
Assuming that the two lowest energy levels ($\hbar \omega_0$ and $\hbar \omega_1$)  are well separated by the higher energy levels and considering  the system of dipoles interacting with a field mode  of frequency $\omega_c \sim \omega_x$, where $\omega_x \equiv \omega_{1,0} $ (here $\omega_{i,j}\equiv \omega_i - \omega_j$), we can truncate the Hibert space of each dipole, by considering as basis only the two lowest energy levels. In this case, each dipole can be modeled as a pseudo-spin, and the Hamiltonian describing the system of $N$ dipoles, in the absence of interaction with the electromagnetic field, can be written in
terms of collective angular momentum operators $\hat J_\alpha = (1/2) \sum_{i =1 }^{N} \hat \sigma^{(i)}_\alpha$ ($\alpha = x, y, z$) as
\be\label{J}
 \hat {\cal H}^{(N)}_0 = \hat {\Pi}  \hat H^{(N)}_0 \hat \Pi = \hbar \omega_x \left( \hat{J}_z + j \right) \, ,
\ee
where $\hat \sigma^{(i)}_\alpha$ are Pauli matrices and  $j = N/2$, and here $\hat \Pi$ is the operator projecting each effective particle into a two-level space. Notice that, after the projection, the operator $\hat \Pi$ represents the identity operator for the linear space constituted by the tensor product of all the $N$ two-level spaces.
Throughout this article we will use calligraphic
symbols (as, for example, $\hat {\cal H}^{(N)}_{0}$) to indicate quantum operators in truncated Hilbert spaces.
Notice that the ground state of the system corresponds to all the spins in their ground state: $| j, j_z = -j \rangle$, and it is an eigenstate of $\hat {\cal H}^{(N)}_0$ with eigenenergy equal to zero.
When all the dipoles are in their excited state, the corresponding collective state $| j, j_z = j \rangle$ has energy $\hbar \omega_x N$.

\subsection{The quantum Dicke model in the Coulomb gauge}

By applying the minimal coupling replacement, the Hamiltonian for the system constituted by $N$ dipoles and a single-mode electromagnetic resonator in the Coulomb gauge can be written  as
\be \label{Hcgfull}
\hat H^{(N)}_{\rm cg} = \sum_{i=1}^{N} \left[\frac{ (\hat p_i - q \hat A)^2}{2 m} + V(x_i)
\right] + \hat H_c\, ,
\ee
where $\hat H_c =  \hbar \omega_c \hat a^\dag \hat a$ is the bare photonic Hamiltonian including a single mode  with resonance frequency $\omega_c$
and annihilation (creation) operator $\hat a$ ($\hat a^\dag$), and $\hat A = A_0(\hat a + \hat a^\dag )$ is the vector potential along the $x$
direction with a zero-point amplitude $A_0$. Notice that the vector potential has been assumed to be constant in the spatial region where the dipoles are present. This approximation can be relaxed, even maintaining the dipole approximation.

It has been shown \cite{Savasta1995,DiStefano2019} that the minimal coupling replacement $\hat p  \to \hat p - q \hat A$ determining \eqref{Hcgfull} can also be implemented by applying to the matter system Hamiltonian  the following unitary transformation
\be\label{SSC}
\hat H^{(N)}_{\rm cg} = \hat U_N \hat H^{(N)}_0 \hat U_N^\dagger + \hat H_c\, ,
\ee
where
\be
\hat U_N = \exp{\left (i \frac{q}{\hbar} \hat A   \sum_{i=1}^N x_i  \right)}\, .
\ee
By expanding the kinetic terms, \eqref{Hcgfull} can be written as the sum of three contributions:
\be
\hat H^{(N)}_{\rm cg} = \hat H^{(N)}_0 + \hat H_c + \hat V^{(N)}_{\rm cg}\, , 
\ee
where  $ \hat V_{cg} = \hat V_{Ap} + \hat V_{D}$ describes the interaction terms
\be
\hat V^{(N)}_{Ap} = \hat A \sum_{i = i}^N \frac{\hat p_i}{m}\, ,
\ee 
and
\be
 \hat V^{(N)}_{D} = N \frac{q^2}{2m} \hat A^2 = D (\hat a + \hat a^\dag)^2\, ,
\ee
where  $D = N A^2_0 q^2/(2m)$.
Using the Thomas–Reiche–Kuhn (TRK) sum rule \cite{Sakurai1994}, the coefficient in the diamagnetic term can be written as $q^2 /{2 m}= \sum_k \omega_{k,j} |d_{k,j}|^2/ \hbar$, where $d_{k,j} =  \langle \psi_k| q x| \psi_j \rangle$ are the dipole matrix elements between two energy eigenstates of the effective particle, that in the following we assume to be real quantities. 
The TRK sum rule has a precise physical meaning, since it expresses the fact that the paramagnetic and diamagnetic contributions to the physical current-current response function cancel in the uniform static limit, which is a consequence of gauge invariance \cite{Pines1966,Giuliani2005,Andolina2019}. The physical current operator, corresponding to the Hamiltonian in \eqref{Hcgfull}, is
\be
\hat {J}_{\rm phys} =  \frac{\delta \hat H_{\rm cg}}{\delta \hat A} 
=  q\sum^N_{i=1} \frac{\hat p_i}{m} + N \frac{q^2}{m} \hat A\, ,
\ee
and the corresponding
current-current response function in the uniform static limit is proportional to \cite{Andolina2019}
\be\label{TRK}
 -2 N \sum_k \omega_{k,j} |d_{k,j}|^2 + N\frac{\hbar q^2 }{m} = 0\,  .
\ee
This relationship expresses  the fact that the paramagnetic (first term on the left hand side) and diamagnetic (second term on the left hand side) contributions to the physical current-current response function cancel out in the uniform and static limit \cite{Andolina2019}. It is interesting to observe that the TRK sum rule remains valid even in the presence of interatomic potentials \cite{Andolina2019}. Very recently, a TRK sum rule for the electromagnetic field coordinates, which holds even in the presence of USC interaction with a matter system, has been proposed \cite{Savasta2020}.

Defining the adimensional coupling strengths $\eta_k = A_0 d_{k,0} /\hbar$, the diamagnetic coefficient can be written as 
\be \label{D}
D = N \hbar  \sum_k  \omega_{k,0}\,  \eta_k^2\, .
\ee

The standard Dicke Hamiltonian in the Coulomb gauge can be obtained from \eqref{Hcgfull}  truncating the Hilbert space of each dipole to include only two energy levels:
\be\label{Hcgs}
\mathcal{H}^{'(N)}_{\rm cg}=  \hat \Pi  \hat H^{(N)}_{\rm cg} \hat \Pi = \omega_c \hat{a}^\dag \hat{a} + \hbar  \omega_x (\hat{J}_z + j) +2  \hbar \omega_x \eta (\hat{a}^\dag + \hat{a})\hat{J}_y + j \frac{q^2 \mathcal{A}_0^2}{m} (\hat{a}^\dag + \hat{a})^2\, ,
\ee
where $\eta \equiv \eta_1= A_0 d_{1,0}/ \hbar$, and the relation $i \hbar p_i/m = [x_i, H^{(i)}_0]$ has been used. 

It has been shown that the two-level truncation for the effective particles ruins the gauge invariance \cite{DeBern2018}. In particular, it has been argued that the Coulomb-gauge Hamiltonian in \eqref{Hcgs} is not related by a unitary transformation (hence it is not gauge equivalent) to the corresponding Hamiltonian in the dipole gauge. Closely related developments have been presented in Refs.~\cite{DeBernardis2018,Stokes2019,DiStefano2019}.
We will discuss this issue in detail below. Here we limit to show that the Hamiltonian in  \eqref{Hcgs} does not satisfy  the gauge principle  and how to solve this problem  following Ref.~\cite{DiStefano2019}.  This Hamiltonian can be obtained, projecting in two-level spaces the full Hamiltonian in  \eqref{Hcgfull}.
Using \eqref{SSC}:
\be
\hat {\cal H}^{'(N)}_{\rm cg}  = \hat \Pi \hat U_N \sum_i \left[\frac{\hat p_i^2}{2m}+ V(x_i) \right] \hat U_N^\dag \hat \Pi + \hbar \omega_c \hat a^\dag \hat a\,  .
\ee
By applying the unitary operator to the kinetic and potential terms separately, observing that $[V(x_i), \hat U_N ]=0$, we obtain
\be\label{Hpcg1}
\hat {\cal H}^{'(N)}_{\rm cg}  = \hat \Pi  \sum_i \frac{(\hat p_i - q \hat A)^2}{2m} \hat \Pi
+ \hat \Pi \sum_i V(x_i)  \hat \Pi + \hbar \omega_c \hat a^\dag \hat a\,  .
\ee
It has been shown that truncating the Hilbert space transforms a local operator like $V(x_i)$ into a nonlocal one which can be expressed as a function of both position and momentum \cite{Starace1971}: $\hat \Pi V(x_i) \hat \Pi = W(x_i, \hat p_i)$. Therefore, the Hamiltonian in \eqref{Hpcg1} contains operators [$W(x_i, \hat p_i)$] depending also on the particle momenta, where the minimal coupling replacement, prescribed by the gauge principle, has not been applied.

\begin{figure}[h]  
	\centering
	\includegraphics[scale=0.4]{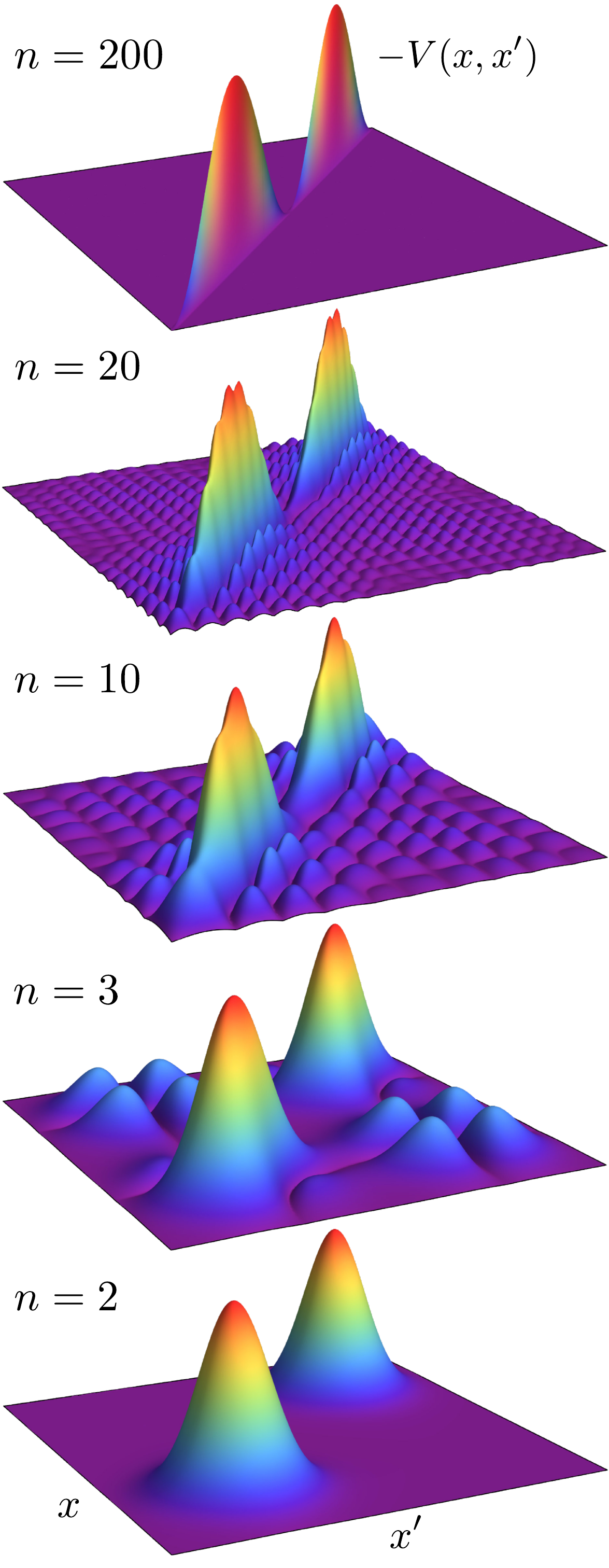}  
	\caption{Example of nonlocal potentials $V(x,x')$ originating from a local potential $V(x)$ (in this case a double well) after the truncation of the Hilbert space to the lowest $n$  energy levels. Decreasing the number of levels, the degree of nonlocality increases. We considered the potential $V(\tilde x) = E_k \left[ -(\beta/2) {\tilde x}^2 + (\gamma/4) {\tilde x}^4\right]$, where $\tilde x$ is an adimensional coordinate \cite{DeBernardis2018},
	$\beta = 3.95$ and
		$\gamma =2.08$
		are adimensional coefficients, and $E_k$ is the kinetic energy coefficient: $\hat H_0  = E_k\, \hat {\tilde p}^2/2 + V(\tilde x)$.  Notice that only adimensional quantities, as a function of  adimensional quantities, have been plotted and the three axes have been omitted.}
	\label{fig2}
\end{figure}

In particular, we observe that for a local potential, we have $\langle {x}'|V|{x}\rangle = V(x)  \delta({x}-{x}')$. By using the closure relation, it can be expressed as $V(x,x') = \sum_{n,n'} V_{n.n'} \psi_n(x) \psi^*_{n'}(x')$, where $\psi_n(x)  = \langle x| \psi_n \rangle$  and $\{|\psi_n \rangle\}$ constitute a complete orthonormal  basis. Notice that the Dirac delta function can be reconstructed only by keeping all the infinite vectors of the basis. Hence, any truncation of the complete basis can
transform a local potential into a non-local one.
The action of the resulting nonlocal potential on a generic state $|\psi\rangle$ in the position representation is
\be\label{nl}
\langle{x}|V|\psi\rangle=\int d{x'}\langle {x}|V|{x'}\rangle \langle { x}'|\psi\rangle=
\int d{x'}V(x,x') \rangle\psi( { x}')\,.
\ee
Using the translation operator property, $\langle {x}|{\hat T}_{a} |\psi\rangle=\exp[i({a}-{x})\hat p]\psi( {x})$, we obtain from \eqref{nl}
\be
\langle{x}|V|\psi\rangle=
\int d{x'} V(x,x')  e^{i({a}-{x})\hat p} \psi( {x}) = V(x, \hat p) \psi (x)\, .
\ee
As an example, \figref{fig2} shows as a local potential $V(x)$ (in this case a double-well potential) evolves into a nonlocal one when increasing the truncation of the Hilbert space. Here $n$ indicates the number of energy states included in the projection operator, starting from the ground state. 

A formulation preserving the gauge principle  can be obtained replacing in \eqref{Hpcg1} the terms  $\hat \Pi V(x_i) \hat \Pi = W(x_i, \hat p_i)$ with  $\hat \Pi W(x_i, \hat p_i - q \hat A) \hat \Pi$. Hence, this problem, arising from the truncation of the Hilbert space of the matter system, can be overcome by first applying to the matter system Hamiltonian (in the absence of interaction) $\hat H_0$ the projection operator $\hat \Pi$, and then the unitary operator $\hat U_N$ as follows: $\hat H^{(N)}_0 \to \hat \Pi \hat H^{(N)}_0 \hat \Pi \to \hat U_N \hat \Pi  \hat H^{(N)}_0 \hat \Pi  \hat U_N^\dag$. Finally, if one asks that the resulting Hamiltonian be within the truncated Hilbert space, one has to  finally project: $\hat U_N \hat \Pi \hat H^{(N)}_0 \hat \Pi  \hat U_N^\dag \to \hat \Pi \hat U_N \hat \Pi \hat H^{(N)}_0 \hat \Pi \hat U_N^\dag \hat \Pi$. This method is not limited to truncated two-level spaces but can be applied to  any truncated Hilbert space to produce light-matter interaction Hamiltonians satisfying the gauge principle. Applying this procedure, we obtain
\be
\label{12}
\hat {\cal H}^{(N)}_{\rm cg}  = \hat {\cal U}_N  \hat{\cal  H}^{(N)}_0  \hat {\cal U}_N^\dag + \hbar \omega_c \hat a^\dag \hat a\, ,
\ee
where $ \hat {\cal U}_N  = \hat \Pi \hat U_N \hat \Pi$.
Using repeatedly the properties of  identity operators $\hat \Pi = \hat \Pi^2$, we obtain
\be
\hat {\cal U}_N = \exp{\left[ 2 i \eta (\hat a + \hat a^\dag) \hat J_x \right]}\, .
\ee
The unitary transformation $ \hat {\cal U}_N  \hat{\cal  H}_0  \hat {\cal U}_N^\dag$ describes the rotation of the system of pseudospins around the $x$ axis  by an angle $\hat \phi =  2 \eta (\hat a + \hat a^\dag)$. The resulting Hamiltonian is 
\be \label{3}
\hat {\cal H}^{(N)}_{\rm cg} =  \hbar  \omega_c \hat{a}^\dag \hat{a}+ \hbar j \omega_x  +  \hbar \omega_x \left\lbrace \hat{J}_z \cos \left[ 2 \eta(\hat{a}^\dag + \hat{a}) \right]+  \hat{J}_y \sin \left[ 2 \eta(\hat{a}^\dag + \hat{a}) \right] \right\rbrace\, .
\ee
This result shows that the occurrence of a non-local potential, arising from the truncation of the matter system Hilbert space, changes significantly
the structure of the Coulomb-gauge interaction Hamiltonian (see, e.g., \cite{Keeling2007}, for comparison).
The price that one has to pay for preserving the gauge principle in such a truncated space is that the total Hamiltonian contains field operators at all orders, in contrast to the standard Coulomb gauge  Hamiltonian in  \eqref{Hcgs}.

\subsection{The Dicke model in the Dipole  gauge}

The Hamiltonian in the dipole gauge for a collection of $N$ effective particles, $\hat {H}^{(N)}_{\rm dg}$,  corresponds to the Power–Zienau–Woolley Hamiltonian after the dipole approximation.
It can be obtained directly from the Hamiltonian
in the Coulomb gauge with the electric dipole approximation
\eqref{Hcgfull} by means of a gauge transformation, which is also a unitary transformation:  
\be \label{Hdgfull}
\hat H^{(N)}_{\rm dg} = \hat T_N \hat H^{(N)}_{\rm cg} \hat T_N^\dag\,  ,
\ee
where $\hat T_N = \hat U^\dag_N$.
We obtain
\be\label{SSD1}
\hat H^{(N)}_{\rm dg}  =  \hat H^{(N)}_0  + \hat T_N \hat H_cT_N^\dag\, .
\ee
Applying the Baker–Campbell–Hausdorff lemma, we have
\be\label{Hdg}
\hat H^{(N)}_{\rm dg} = \hat H^{(N)}_0 + \hat H_c +i\frac{qA_0}{\hbar} (\hat a^\dag -\hat a) \sum_i x_i + \left(\frac{qA_0}{\hbar}\right)^2\sum_{i,j} x_ix_j
\, .
\ee
The standard Dicke Hamiltonian in the dipole gauge can be obtained from \eqref{Hdg}  truncating the Hilbert space of each dipole to include only two energy levels: 
$\hat {\cal H}^{(N)}_{\rm dg} = \hat \Pi \hat {H}^{(N)}_{\rm dg} \hat \Pi$.
Observing that $  q \hat \Pi \sum_i x_i \hat \Pi = 2 d_{1,0} \hat J_x$, and using the fact that $\hat \Pi$ is the identity operator for the resulting collection of two-level systems, we obtain
\be\label{DickeD}
\hat {\cal H}^{(N)}_{\rm dg} = 
\hbar \omega_c \hat{a}^\dag \hat{a} + \hbar \omega_x (\hat{J}_z +j)+2\,i \hbar \eta \,\omega_c (\hat{a}^\dag - \hat{a}) \hat{J}_x +4 \hbar \,\eta^2 \,\omega_c \,\hat{J}_x^2\, .
\ee
Comparing \eqref{SSC} and \eqref{SSD1} (notice that $\hat T_N = \hat U^\dag_N$), we  observe that while the Coulomb-gauge Hamiltonian can be obtained by applying a unitary transformation to the bare matter Hamiltonian, the dipole-gauge Hamiltonian is obtained by applying the $h.c.$ transformation to the bare photonic Hamiltonian.

We will show in the next subsection that, in contrast to the standard derivation of the Coulomb-gauge Dicke Hamiltonian, the dipole gauge Hamiltonian in \eqref{DickeD} does not violate the gauge principle. This behaviour can be understood by observing that a truncation on the number of modes in the photonic system, as a single-mode description of the resonator, despite determining a loss of spatial locality \cite{Munoz2018}, it does not introduce any spatial nonlocality in the quadratic potential of the single-mode Hamiltonian, since different normal modes are independent and corresponds to different effective particles. On the contrary, truncating the Hilbert space of an individual mode, e.g., considering a few photon system, could produce issues analogous to  those appearing in the Coulomb gauge.

Equation~(\ref{DickeD}) describes the Dicke Hamiltonian in the dipole gauge. It includes a self-polarization term induced by the interaction with the electromagnetic field ($\propto \hat{J}_x^2$). Neglecting it can lead to unphysical results \cite{Schafer2019} and to the loss of gauge invariance. This Hamiltonian slightly differs from that derived in \cite{DeBernardis2018}, where the intra-atom self-polarization terms $\propto x_i^2$ are included in the atomic potentials and give rise to a renormalization  of the atomic transition frequency $\omega_{1,0}$ and of the coupling $\eta$. While the full inclusion of these terms into the qubit Hamiltonian might seem to be the most accurate approach to derive a reduced two-level Hamiltonian, it applies the two-level truncation to the different  terms  of the  light-matter interaction Hamiltonian with a different level of accuracy.
 Specifically, while  the terms $\propto x_i^2$ are included in the atomic potentials before the diagonalization of the atomic Hamiltonian,  the other terms  are taken into account only after the application of the two-level approximation.
Moreover, the resulting self-polarization term $\hat{J}_x^2 = (1/4)\sum_{i,j} \hat \sigma_x^{(i)} \hat \sigma_x^{(j)}$ still includes the intra-atomic contributions ($i=j$), although these determine only a rigid shift of all the energy levels.  In Ref.~ \cite{DeBernardis2018} it is shown that, when the coupling strength is quite high, including the intra-atom self-polarization terms in the atom potential before the diagonalization of the full atomic Hamiltonian, can result in less accurate results.

\subsection{Gauge invariance of the Dicke model}
The Dicke Hamiltonian in the dipole gauge in \eqref{DickeD} can also be derived directly applying a gauge (unitary) transformation to the Dicke Hamiltonian in the Coulomb gauge in \eqref{12} [or alternatively in \eqref{3}]:

\be \label{gaugeCD}
\hat {\cal H}^{(N)}_{\rm dg} = \hat {\cal T}_N \hat {\cal H}^{(N)}_{\rm cg} \hat {\cal T}_N^\dag\,  ,
\ee
where $\hat {\cal T}_N = \hat {\cal U}^\dag_N$.
Equation~(\ref{gaugeCD}) demonstrates that the two formulations of the Dicke model $\hat {\cal H}^{(N)}_{\rm cg}$ and $\hat {\cal H}^{(N)}_{\rm dg}$ are related by a gauge transformation. Such a relation is not fulfilled if $\hat {\cal H}^{(N)}_{\rm cg}$ is replaced by
$\hat {\cal H}^{'(N)}_{\rm cg}$.

\section{The Dicke Model in the $N \to \infty$ limit}

The starting point for our analysis in the thermodynamic limit is the Holstein-Primakoff representation \cite{Holstein1940} of the angular momentum operators
$\hat J_z = \hat b^\dag \hat b - j$, $\hat J_+ = \hat b^\dag \sqrt{2j -\hat b^\dag \hat b}$, $\hat J_- = \hat J^\dag_+$ [notice that $\hat J_\pm = (\hat J_x \pm  i \hat J_y)/2$]. Here $\hat b$ and $\hat b^\dag$ are bosonic operators. 
This allows to obtain effective Hamiltonians that are exact in the standard thermodynamic limit $N \to \infty$, $\eta \to 0$, with $\eta \sqrt{N} \to  \lambda$ remaining a finite quantity.

We proceed in the thermodynamic limit by replacing the angular momentum operators introduced  in the previous section by using the Holstein-Primakoff  representation, expanding the square roots and finally neglecting terms with powers of $j$ in the denominator, since these go to zero in the considered limit \cite{Cortese2017}. We can start from the Hamiltonian of the collective spin system in the absence of interaction with the electromagnetic field in \eqref{J}. We obtain
\be\label{H0bose}
\hat {\cal H}_0 = \hbar \omega_x \hat b^\dag \hat b\, .
\ee

\subsection{Dipole gauge}
Applying the Holstein-Primakoff  representation to \eqref{DickeD}, performing the thermodynamic limit ($N \to \infty$ and $\eta \sqrt{N} \to \lambda$), we obtain
\be\label{DickeDbose}
\hat {\cal H}_{\rm dg} = 
\hbar \omega_c \hat{a}^\dag \hat{a} + \hbar \omega_x \hat b^\dag \hat b +i \hbar \lambda \,\omega_c (\hat{a}^\dag - \hat{a}) (\hat b + \hat b^\dag ) +\hbar \omega_c\, \lambda^2 \,(\hat b + \hat b^\dag)^2\,  .
\ee

\subsection{Coulomb gauge}

In contrast to the Dicke Hamiltonians in the dipole gauge $\hat {\cal H}^{(N)}_{\rm dg}$, and in the standard Coulomb gauge  $\hat {\cal H}^{'(N)}_{\rm cg}$, the correct Coulomb gauge Dicke Hamiltonian $\hat {\cal H}^{(N)}_{\rm cg}$
contains field operators at all orders. At a first sight, this feature prevents the possibility to obtain a harmonic Dicke Hamiltonian in the thermodynamic limit  as obtained from $\hat {\cal H}^{(N)}_{\rm dg}$. Hence, the thermodynamic limit, apparently, would destroy  gauge invariance. Actually, as we are going to show, this is not the case.

Starting from \eqref{3}, performing a series expansion of $ \cos \left[ 2 \eta(\hat{a}^\dag + \hat{a}) \right] $ and  $ \sin \left[ 2 \eta(\hat{a}^\dag + \hat{a}) \right] $, we obtain
\begin{align}
\hat {\cal H}^{(N)}_{\rm cg} &= \hbar \omega_c \hat{a}^\dag \hat{a}+ \hbar\frac{N  \omega_x}{2}+  \hbar \omega_x  \left( \hat{b}^\dag \hat{b}-N/2\right)  \left[ 1- 2 \eta^2 (\hat{a}^\dag + \hat{a})^2 + \mathcal{O}(\eta^4) \right]& \nonumber\\ 
&-i \hbar \omega_x\frac{\sqrt{N}}{2}  (\hat{b}^\dag - \hat{b})  \left[  2 \eta(\hat{a}^\dag + \hat{a}) + \mathcal{O}(\eta^3) \right]\, . &
\end{align}
In the thermodynamic limit ($N \to \infty$, $\sqrt{N} \eta \to \lambda$), only terms up to the second order in $\eta$ remain different from zero, and we finally obtain
\be \label{7}
\mathcal{H}_{\rm cg}= \hbar   \omega_c \hat{a}^\dag \hat{a} +  \hbar \omega_x \hat{b}^\dag \hat{b}-  i  \hbar \omega_x \lambda\,(\hat{b}^\dag - \hat{b})(\hat{a}^\dag + \hat{a})+  \hbar {\cal D} (\hat{a}^\dag + \hat{a})^2\, ,
\ee
where we defined ${\cal D} =  \omega_x \lambda^2$.
As a result, also the correct Coulomb gauge Hamiltonian $\mathcal{H}^{(N)}_{\rm cg}$ [\eqref{3}],  reduces to an Hamiltonian which describes an harmonic system constituted by two interacting harmonic oscillators, like the dipole gauge Hamiltonian.

In the same limit,  the standard Coulomb gauge Hamiltonian $\mathcal{H}^{'(N)}_{\rm cg}$, not satisfying the gauge principle becomes
\be \label{8}
\mathcal{H}'_{\rm cg}= \hbar   \omega_c \hat{a}^\dag \hat{a} +  \hbar \omega_x \hat{b}^\dag \hat{b}-  i  \hbar \omega_x \lambda\,(\hat{b}^\dag - \hat{b})(\hat{a}^\dag + \hat{a})+  \hbar {\cal D}'  (\hat{a}^\dag + \hat{a})^2\, ,
\ee
where we used \eqref{D}, and defined ${\cal D}' =  \sum_k \omega_{k,0}\lambda_k^2 = D/\hbar$.
$\hat {\cal H}'_{\rm cg}$ in \eqref{8} is very similar to $\hat {\cal H}_{\rm cg}$ in \eqref{7}. They only differ for the diamagnetic coefficient multiplying the term $(\hat{a}^\dag + \hat{a})^2$. While the coefficient in \eqref{8} (${\cal D}'$) contains a sum over all the allowed transitions from the ground state, the one in \eqref{7} (${\cal D}  < {\cal D}'$), more consistently, contains only the contribution from the single two-level transition considered in the two-level approximation leading to the Dicke model. As we will show in the next subsection, this difference determines the loss or the preservation of gauge invariance. Moreover, it has been observed that the value of the diamagnetic coefficient with respect to $\omega_x \lambda^2$ can prevent or allow a superradiant phase transition in Dicke models \cite{Nataf2010a}.

It is interesting and reassuring that also after the  truncation of the Hilbert space of the atomic ensemble, using \eqref{7}, the paramagnetic and diamagnetic contributions to the physical current-current response function \cite{Pines1966,Giuliani2005,Andolina2019} still cancel in the uniform static limit.  In particular, in the present case, it is proportional to

\be\label{TRK}
- \frac{(\omega_x \lambda)^2}{\omega_x}+ {\cal D} = 0\,  .
\ee
This does not occur using the Hamiltonain in \eqref{8}:
\be\label{TRK'}
- \frac{(\omega_x \lambda)^2}{\omega_x}+ {\cal D}'  \neq 0\,  .
\ee

\subsection{Gauge invariance}
\label{GI}
In order to demonstrate that $\hat {\cal H}_{\rm cg}$ and $\hat {\cal H}_{\rm dg}$ are related by a unitary (gauge) transformation and hence display the same spectrum of eigenergies, we start applying the 
Holstein-Primakoff  representation to the unitary operator which implements the minimal coupling replacement in \eqref{12}, as well as the gauge transformation of the Dicke model [see \eqref{gaugeCD}].
Taking the standard limits  ($N \to \infty$, with  $\sqrt{N} \eta = \lambda$ finite), we obtain
\be\label{Uinf}
\hat {\cal U}_N  \to  \hat {\cal U} = \exp{\left[  i \lambda  (\hat a + \hat a^\dag)( \hat b + \hat b^\dagger) \right]}\, .
\ee
The Dicke Hamiltonian in the Coulomb gauge $\hat {\cal H}_{\rm cg}$ can be readily obtained by applying the generalized minimal coupling replacement using \eqref{H0bose} and \eqref{Uinf}:
\be
\label{12inf}
\hat {\cal H}_{\rm cg}  = \hat {\cal U}  \hat{\cal  H}_0\,  \hat {\cal U}^\dag + \hbar \omega_c \hat a^\dag \hat a\, .
\ee

This approach is particularly interesting, since it provides a recipe to obtain the correct Coulomb-gauge light-matter interaction Hamiltonian starting from the knowledge of the unperturbed Hamiltonian of a bosonic excitation
$ \hat{\cal  H}_0$ and its associated polarization operator, which in this case is $\hat p = \sqrt{N} d_{1,0} (\hat b + \hat b^\dag)$.
 Notice that the unitary operator in \eqref{12inf} can be expressed as $\hat {\cal U} = \exp{\left(i \hat A \hat p /\hbar \right) }$. Thus,  within this approach, it is not necessary to start explicitly considering a collection of effective two-level atoms, but it is sufficient to start from a bosonic Hamiltonian for the bare matter system and then to use the generalized minimal coupling replacement in \eqref{12inf}. We will discuss further this point and its connection with  the Hopfield model in the next section.

Applying  to  $\hat {\cal H}_{\rm cg}$ the unitary transformation $\hat {\cal T} \hat {\cal H}_{\rm cg} \hat {\cal T}^\dag$, where $\hat {\cal T} = \hat {\cal U}^\dag$, the corresponding Hamiltonian in the dipole gauge in \eqref{DickeDbose} is easily recovered: 
\be \label{gaugeCDinf}
\hat {\cal T} \hat {\cal H}_{\rm cg} \hat {\cal T}^\dag =  \hat {\cal H}_{\rm dg} \, .
\ee
Equation~(\ref{gaugeCDinf}) demonstrates that $ \hat {\cal H}_{\rm dg}$ and  $\hat {\cal H}_{\rm cg}$ are related by a unitary transformation as required by gauge invariance, hence they will display the same eigenvalues.
In contrast, $\hat {\cal H}'_{\rm cg}$ is \textit{not} related to $ \hat {\cal H}_{\rm dg}$ by a unitary  transformation and thus it will display \textit{different} energy levels. 

We now provide a direct check of the breakdown of gauge invariance of the Dicke model as described by the standard Hamiltonian in the Coulomb gauge \eqref{8}. Specifically, we compare  the resonance frequencies of the two collective polariton modes obtained by diagonalizing (using Bogoliubov-Hopfield transformations) the Hamiltonians \eqref{DickeDbose}, \eqref{7}, and  \eqref{8}. For the polariton frequencies, resulting from the diagonalization of  \eqref{DickeDbose}, we obtain
\be\label{wD}
\omega^2_{{\rm dg} \pm} = \frac{1}{2} \left[ \tilde \omega^2_x + \omega^2_c  \pm
\sqrt{(\tilde \omega^2_x - \omega^2_c)^2 + 4 \lambda^2 \omega_x \omega_c}
\right] \, ,
\ee
where  $\tilde \omega_x = \sqrt{\omega_x (\omega_x + 4 \lambda^2 / \omega_c)}$.

Diagonalizing the Hamiltonian in \eqref{7} results into the polariton frequencies
\be\label{wC}
\omega^2_{{\rm cg} \pm} =\frac{1}{2}\left[\tilde{\omega}_c^2 +\omega_x^2\pm \sqrt{(\tilde{\omega}_c^2 +\omega_x^2)^2 - 4 \,\omega_c^2 \omega_x^2 } \right] ,
\ee
with $ \tilde{\omega}_c = \sqrt{\omega_c(\omega_c+ 4 \mathcal{D})} $.

 The polariton frequencies $\omega'_{{\rm cg} \pm}$ resulting from the diagonalization of the standard Coulomb-gauge Dicke Hamiltonian in \eqref{8} can be obtained from \eqref{wC} after the replacement ${\cal D} \to {\cal D'}$.
 
The unitary gauge transformation in \eqref{gaugeCDinf} implies that $\omega_{{\rm dg} \pm} = \omega_{{\rm cg} \pm}$. This relation can be explicitly shown after some algebric manipulation.
On the contrary, the polariton frequencies obtained from $\hat {\cal H}'_{\rm cg}$ are different:
\[
\omega'_{{\rm cg} \pm} \neq \omega_{{\rm cg} \pm} = \omega_{{\rm dg} \pm}\, .
\]
 \begin{figure}[H]  
 	\centering
 	\includegraphics[scale=0.53]{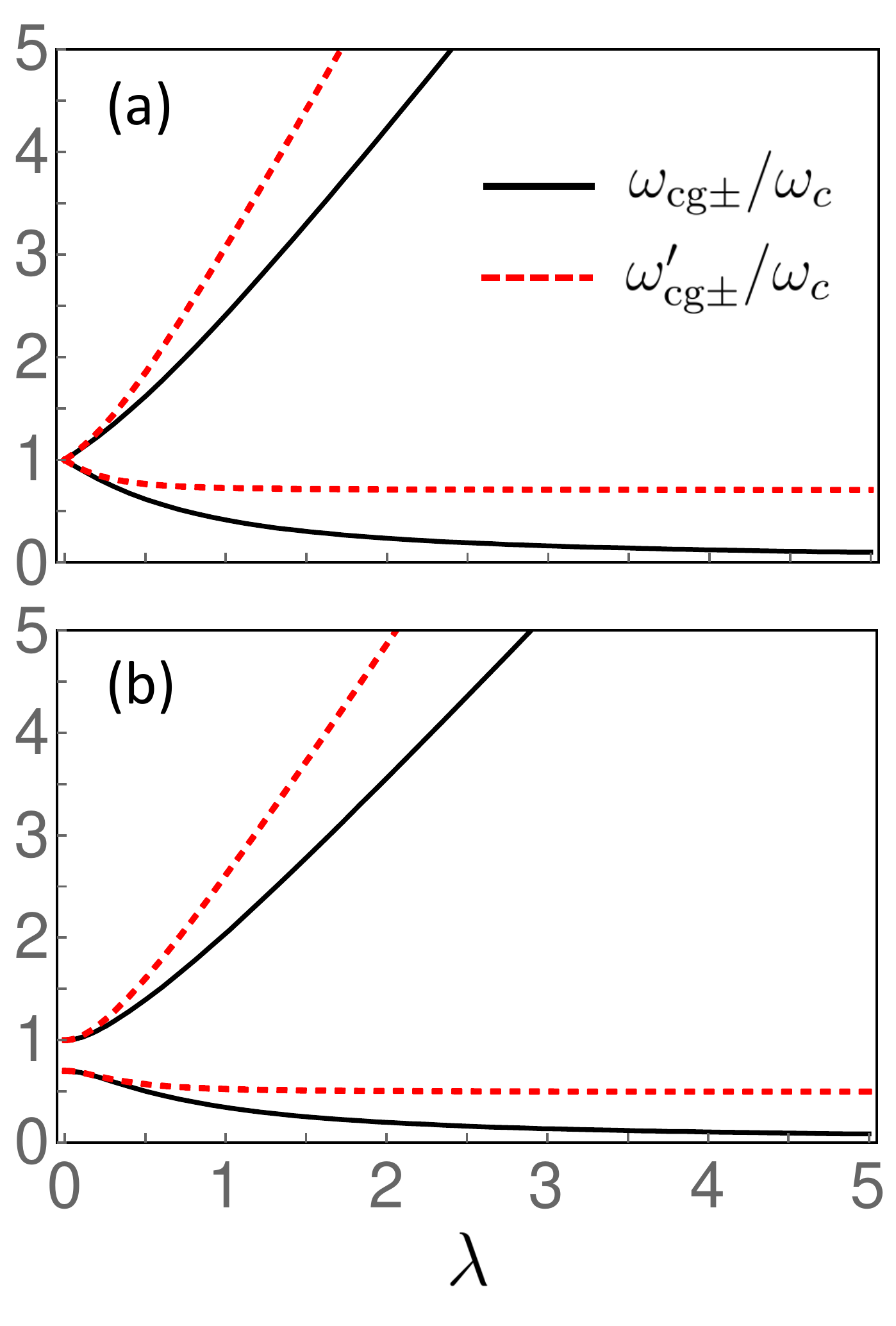}  
 	\caption{The frequencies $\omega_{{\rm cg}\pm} = \omega_{{\rm dg}\pm}$ and $\omega'_{{\rm cg}\pm}$ of the two polariton modes, obtained diagonalizing the Dicke model, in the limit $ N \to \infty$ as a function of the normalized  coupling
 		strength $\lambda$ for (a) the resonant case ($\omega_c=\omega_x $); (b) for the detuned case with $\omega_x = 0.8 \omega_c$.} 
 	\label{fig3}
 \end{figure}
Figure~(\ref{fig3}) displays $ \omega_{{\rm cg} \pm}/ \omega_c = \omega_{{\rm dg} \pm}/\omega_c$
and $ \omega'_{{\rm cg} \pm}/ \omega_c$ as a function of $\lambda$, for ${\cal D}' = 2 {\cal D}$. 
The choice of  $\alpha \equiv {\cal D}'/{\cal D} $ depends on the specific system. Here we used the reasonable value $\alpha =2$.

The differences are relevant, starting from normalized coupling strengths $\lambda \sim 0.4$. 
Hence, we can conclude that for coupling strengths $\lambda \gtrsim 0.4$ the standard Coulomb-gauge Dicke Hamiltonian (in the thermodynamic limit) provides significantly wrong polariton frequencies in agreement with the results in Ref.~\cite{DeBernardis2018}.

\section{Gauge invariance of the Hopfield model}

The Hopfield model provides a full quantum description of the interaction between the electromagnetic field and  a dielectric which is described by a harmonic polarization density.
The original treatment considers a 3D uniform  dielectric with a single resonance frequency describing  dispersionless collective excitations. This exactly solvable model was initially applied to the case of excitonic polaritons, then, it has been applied and/or generalized to describe a great variety of systems with different dimensionalities and degrees of freedom, including quantum well \cite{Savasta1996} and cavity polaritons \cite{Savona1994}, phonon-polaritons \cite{Gubbin2016a,Sentef2018}, and plasmonic  nanoparticle crystals \cite{Lamowski2018}. A generalized Hopfield model  for inhomogeneous and dispersive media has been proposed \cite{Gubbin2016}.
Here we analyze the original model, its gauge properties and its connection with the Dicke model in the thermodynamic limit.

The field operators are given in terms of the bosonic  photonic operators $\hat a_{{\bf k}, \lambda}$ and 
the bosonic operators $\hat b_{{\bf k}, \lambda}$ describing the destruction of the polarization quanta by
\bea\label{fieldos}
\hat A ({\bf r}) &=& \sum_{{\bf k}, \lambda} A_k^{(0)} {\bf e}_{{\bf k}, \lambda} \left( \hat a_{{\bf k}, \lambda} + \hat a^\dagger_{-{\bf k}, \lambda} \right) e^{i {\bf k} \cdot {\bf r} }\, ,\nonumber \\
	\hat {\bf P}({\bf r}) &=& P^{(0)}\sum_{{\bf k}, \lambda} {\bf e}_{{\bf k}, \lambda} \left( \hat b_{{\bf k}, \lambda} + \hat b^\dagger_{-{\bf k}, \lambda} \right) e^{i {\bf k} \cdot {\bf r} }\, ,
\eea
where ${\bf k}$ is the wavevector, $\lambda$ labels the two transverse polarizations, ${\bf e}_{{\bf k}, \lambda}$ are the polarization unit vectors,  and we have defined $A_k^{(0)}=\sqrt{{\hbar}/{(2\epsilon_0 V \omega_k)}}$ and $ P^{(0)}=\sqrt{{\hbar \omega_0 \beta}/{(2 V )}}$. Here, $V$ is the quantization volume, $\omega_k$ and $\omega_0$ are the bare resonance frequencies of the photonic modes and of the matter system waves respectively, and $\beta$ is the polarizability \cite{Hopfield1958}.

The Hopfield Hamiltonian in the Coulomb gauge can be written as
\bea\label{HChop}
\hat H^{\rm Hop}_{\rm cg} &=&  \hbar  \sum_{{\bf k}, \lambda}\omega_k \hat a^\dag_{{\bf k}, \lambda} \hat a_{{\bf k}, \lambda} + \hbar \omega_0 \sum_{{\bf k}, \lambda} \hat b^\dag_{{\bf k}, \lambda} \hat b_{{\bf k}, \lambda} \nonumber \\
&+&i \hbar \omega_0 \sum_{{\bf k}, \lambda}\Lambda_k \left( \hat a_{{\bf k}, \lambda} + \hat a^\dagger_{-{\bf k}, \lambda} \right) \left( \hat b_{{\bf k}, \lambda} - \hat b^\dagger_{-{\bf k}, \lambda} \right) 
+ \hbar \omega_0 \sum_{{\bf k}, \lambda} \Lambda^2_k \left( \hat a_{{\bf k}, \lambda} + \hat a^\dagger_{-{\bf k}, \lambda} \right)^2\, ,
\eea
where $\Lambda_k = V A^{(0)}_k P^{(0)}  /\hbar$.

It is interesting to observe that this equation can be written  in the compact form
\be\label{GMCR}
\hat H^{\rm Hop}_{\rm cg} = \hbar  \sum_{{\bf k}, \lambda} \omega_k\hat a^\dag_{{\bf k}, \lambda} \hat a_{{\bf k}, \lambda} +  \hat U_{\rm Hop}   \left(  \hbar \omega_0\sum_{{\bf k}, \lambda} \hat b^\dag_{{\bf k}, \lambda} \hat b_{{\bf k}, \lambda}\right) \hat U^\dag_{\rm Hop}\, ,
\ee
where 
\be
 \hat U_{\rm Hop}   = \exp{\left[ i \sum_{{\bf k}, \lambda}\Lambda_k \left( \hat a_{{\bf k}, \lambda} + \hat a^\dagger_{-{\bf k}, \lambda} \right) \left( \hat b_{{\bf k}, \lambda} - \hat b^\dagger_{-{\bf k}, \lambda} \right) \right]}\, .
\ee
We observe that this unitary operator coincides with the hermitian conjugate of the operator describing the Coulomb $\to$ dipole gauge transformation in  a system with a polarization density operator given by \eqref{fieldos}: 
\be\label{UT}
\hat U_{\rm Hop} = \hat T^\dag_{\rm Hop}\, ,
\ee
where
\be\label{T}
\hat T_{\rm Hop} = \exp{\left[\frac{i}{\hbar} \int d {\bf r} \hat A ({\bf r}) \cdot \hat P ({\bf r})\right]}\, .
\ee

This relationship implies that the Hopfield Hamiltonian in the dipole gauge can be easily obtained: 
\be\label{GH}
\hat H^{\rm Hop}_{\rm dg} =  \hat T_{\rm Hop}  \hat H^{\rm Hop}_{\rm cg} \hat T^\dag_{\rm Hop}
= 
\hat T _{\rm Hop}   \left( \hbar \sum_{{\bf k}, \lambda} \omega_k \hat a^\dag_{{\bf k}, \lambda} \hat a_{{\bf k}, \lambda} \right) \hat T^\dag_{\rm Hop} +    \hbar \omega_0\sum_{{\bf k}, \lambda} \hat b^\dag_{{\bf k}, \lambda} \hat b_{{\bf k}, \lambda}\, .
\ee
After simple algebra, we obtain
\bea\label{Hdhop}
\hat H^{\rm Hop}_{\rm dg} &=&  \hbar  \sum_{{\bf k}, \lambda}\omega_k \hat a^\dag_{{\bf k}, \lambda} \hat a_{{\bf k}, \lambda} + \hbar \omega_0 \sum_{{\bf k}, \lambda} \hat b^\dag_{{\bf k}, \lambda} \hat b_{{\bf k}, \lambda} \nonumber \\
&-&i \hbar  \sum_{{\bf k}, \lambda}\omega_k\Lambda_k \left( \hat a_{{\bf k}, \lambda} - \hat a^\dagger_{-{\bf k}, \lambda} \right) \left( \hat b_{{\bf k}, \lambda} + \hat b^\dagger_{-{\bf k}, \lambda} \right) 
+ \hbar\sum_{{\bf k}, \lambda}\omega_k \Lambda^2_k \left( \hat b_{{\bf k}, \lambda} + \hat b^\dagger_{-{\bf k}, \lambda} \right)^2\, .
\eea
Equation~(\ref{GH}) demonstrates that \eqref{HChop} and \eqref{Hdhop} are related by a unitary (gauge) transformation and hence displays the same energy spectrum. The compact forms in \eqref{GMCR} and \eqref{GH} are manifestly gauge related. Moreover being manifestly related by a unitary transformation, they provie the same energy spectra and the same matrix elements of physical observables. Of course, both the corresponding operators and the vector states have to be transformed accordingly, when changing from one gauge to the other. 
If needed, a continuous set of gauge transformations depending by one parameter can be considered. It is sufficient to, e.g., to start from the Hamiltonian in the Coulomb gauge and then considering unitary transformation using modified unitary operators where the exponent in \eqref{T} is multiplied by such a parameter (see, e.g., \cite{Stokes2019}).

These results open the way to the application of the {\em generalized minimal coupling replacement} [see Eqs.~(\ref{GMCR}), (\ref{UT}), and (\ref{T})] to promptly derive  general gauge-invariant Hopfield Hamiltonians. 
Given a generic polarization operator like that in \eqref{fieldos}, using the unitary operator in \eqref{T}, it is possible to directly obtain the total Hamiltonian in the Coulomb or dipole gauge by applying the corresponding transformation to the bare matter system Hamiltonian [see \eqref{GMCR}] or to the bare photonic Hamiltonian [see \eqref{Hdhop}]. From this point of view, {\em the Dicke model in the dilute regime can be regarded as a particular case of the Hopfield model} where the polarization density operator is  $\hat P = (\sqrt{N}d_{1,0}/V)  (\hat b + \hat b^\dag)$ (see Sect.~\ref{GI}).

\section{Conclusions}
We have investigated the gauge invariance of the Dicke model in the dilute regime. In particular we started from the derivation of  the correct (not violating the gauge principle) Dicke model in the Coulomb gauge for a finite number $N$ of dipoles. After that, using the Holstein-Primakoff transformation, we obtained the Coulomb-gauge Dicke Hamiltonian  in the dilute regime. We demonstrated that it is related by  a gauge (unitary) transformation to the corresponding Hamiltonian in the dipole gauge. Hence the two gauges, as required, provide the same energy spectra, in contrast with the standard Dicke model.
 The standard Dicke Hamiltonian in the Coulomb gauge, and the  one derived here only differ for the diamagnetic coefficient multiplying the term $(\hat{a}^\dag + \hat{a})^2$. This difference determines the loss or the preservation of gauge invariance.

Finally we analyzed the Hopfield model, showing its gauge invariance. We provided a method to derive in a simple way manifestly gauge-invariant  Hopfield models, having  knowledge just of the matter polarization field. These results show that the Dicke model in the dilute regime can be regarded as a particular case of the more general Hopfield model.

{\acknowledgments}

\noindent
We thank Simone De Liberato for useful discussions and suggestions.
F.N. is supported in part by the: 
MURI Center for Dynamic Magneto-Optics via the 
Air Force Office of Scientific Research (AFOSR) (FA9550-14-1-0040), 
Army Research Office (ARO) (Grant No. Grant No. W911NF-18-1-0358), 
Japan Science and Technology Agency (JST) (via the Q-LEAP program, and the CREST Grant No. JPMJCR1676), 
Japan Society for the Promotion of Science (JSPS) (JSPS-RFBR Grant No. 17-52-50023, and 
JSPS-FWO Grant No. VS.059.18N), the RIKEN-AIST Challenge Research Fund, 
the Foundational Questions Institute (FQXi), and the NTT PHI Laboratory. 
S.S. acknowledges the Army Research Office (ARO)
(Grant No. W911NF1910065).

\bibliography{refMEnew}

\begin{thebibliography}{57}%
\makeatletter
\providecommand \@ifxundefined [1]{%
 \@ifx{#1\undefined}
}%
\providecommand \@ifnum [1]{%
 \ifnum #1\expandafter \@firstoftwo
 \else \expandafter \@secondoftwo
 \fi
}%
\providecommand \@ifx [1]{%
 \ifx #1\expandafter \@firstoftwo
 \else \expandafter \@secondoftwo
 \fi
}%
\providecommand \natexlab [1]{#1}%
\providecommand \enquote  [1]{``#1''}%
\providecommand \bibnamefont  [1]{#1}%
\providecommand \bibfnamefont [1]{#1}%
\providecommand \citenamefont [1]{#1}%
\providecommand \href@noop [0]{\@secondoftwo}%
\providecommand \href [0]{\begingroup \@sanitize@url \@href}%
\providecommand \@href[1]{\@@startlink{#1}\@@href}%
\providecommand \@@href[1]{\endgroup#1\@@endlink}%
\providecommand \@sanitize@url [0]{\catcode `\\12\catcode `\$12\catcode
  `\&12\catcode `\#12\catcode `\^12\catcode `\_12\catcode `\%12\relax}%
\providecommand \@@startlink[1]{}%
\providecommand \@@endlink[0]{}%
\providecommand \url  [0]{\begingroup\@sanitize@url \@url }%
\providecommand \@url [1]{\endgroup\@href {#1}{\urlprefix }}%
\providecommand \urlprefix  [0]{URL }%
\providecommand \Eprint [0]{\href }%
\providecommand \doibase [0]{http://dx.doi.org/}%
\providecommand \selectlanguage [0]{\@gobble}%
\providecommand \bibinfo  [0]{\@secondoftwo}%
\providecommand \bibfield  [0]{\@secondoftwo}%
\providecommand \translation [1]{[#1]}%
\providecommand \BibitemOpen [0]{}%
\providecommand \bibitemStop [0]{}%
\providecommand \bibitemNoStop [0]{.\EOS\space}%
\providecommand \EOS [0]{\spacefactor3000\relax}%
\providecommand \BibitemShut  [1]{\csname bibitem#1\endcsname}%
\let\auto@bib@innerbib\@empty
\bibitem [{\citenamefont {Rabi}(1936)}]{Rabi1936}%
  \BibitemOpen
  \bibfield  {author} {\bibinfo {author} {\bibfnamefont {I.~I.}\ \bibnamefont
  {Rabi}},\ }\bibfield  {title} {\enquote {\bibinfo {title} {On the process of
  space quantization},}\ }\href {\doibase 10.1103/PhysRev.49.324} {\bibfield
  {journal} {\bibinfo  {journal} {Phys. Rev.}\ }\textbf {\bibinfo {volume}
  {49}},\ \bibinfo {pages} {324--328} (\bibinfo {year} {1936})}\BibitemShut
  {NoStop}%
\bibitem [{\citenamefont {Rossatto}\ \emph {et~al.}(2017)\citenamefont
  {Rossatto}, \citenamefont {Villas-B\^oas}, \citenamefont {Sanz},\ and\
  \citenamefont {Solano}}]{Rossatto2017}%
  \BibitemOpen
  \bibfield  {author} {\bibinfo {author} {\bibfnamefont {D.~Z.}\ \bibnamefont
  {Rossatto}}, \bibinfo {author} {\bibfnamefont {C.~J.}\ \bibnamefont
  {Villas-B\^oas}}, \bibinfo {author} {\bibfnamefont {M.}~\bibnamefont {Sanz}},
  \ and\ \bibinfo {author} {\bibfnamefont {E.}~\bibnamefont {Solano}},\
  }\bibfield  {title} {\enquote {\bibinfo {title} {Spectral classification of
  coupling regimes in the quantum {R}abi model},}\ }\href {\doibase
  10.1103/PhysRevA.96.013849} {\bibfield  {journal} {\bibinfo  {journal} {Phys.
  Rev. A}\ }\textbf {\bibinfo {volume} {96}},\ \bibinfo {pages} {013849}
  (\bibinfo {year} {2017})}\BibitemShut {NoStop}%
\bibitem [{\citenamefont {{Di Stefano}}\ \emph {et~al.}(2017)\citenamefont {{Di
  Stefano}}, \citenamefont {Stassi}, \citenamefont {Garziano}, \citenamefont
  {Kockum}, \citenamefont {Savasta},\ and\ \citenamefont
  {Nori}}]{DiStefano2017}%
  \BibitemOpen
  \bibfield  {author} {\bibinfo {author} {\bibfnamefont {O.}~\bibnamefont {{Di
  Stefano}}}, \bibinfo {author} {\bibfnamefont {R.}~\bibnamefont {Stassi}},
  \bibinfo {author} {\bibfnamefont {L.}~\bibnamefont {Garziano}}, \bibinfo
  {author} {\bibfnamefont {A.~F.}\ \bibnamefont {Kockum}}, \bibinfo {author}
  {\bibfnamefont {S.}~\bibnamefont {Savasta}}, \ and\ \bibinfo {author}
  {\bibfnamefont {F.}~\bibnamefont {Nori}},\ }\bibfield  {title} {\enquote
  {\bibinfo {title} {{Feynman-diagrams approach to the quantum Rabi model for
  ultrastrong cavity QED: stimulated emission and reabsorption of virtual
  particles dressing a physical excitation}},}\ }\href {\doibase
  10.1088/1367-2630/aa6cd7} {\bibfield  {journal} {\bibinfo  {journal} {New J.
  Phys.}\ }\textbf {\bibinfo {volume} {19}},\ \bibinfo {pages} {053010}
  (\bibinfo {year} {2017})}\BibitemShut {NoStop}%
\bibitem [{\citenamefont {Dicke}(1954)}]{Dicke1954}%
  \BibitemOpen
  \bibfield  {author} {\bibinfo {author} {\bibfnamefont {R.~H.}\ \bibnamefont
  {Dicke}},\ }\bibfield  {title} {\enquote {\bibinfo {title} {Coherence in
  spontaneous radiation processes},}\ }\href {\doibase 10.1103/PhysRev.93.99}
  {\bibfield  {journal} {\bibinfo  {journal} {Phys. Rev.}\ }\textbf {\bibinfo
  {volume} {93}},\ \bibinfo {pages} {99--110} (\bibinfo {year}
  {1954})}\BibitemShut {NoStop}%
\bibitem [{\citenamefont {Brandes}(2005)}]{Brandes2005}%
  \BibitemOpen
  \bibfield  {author} {\bibinfo {author} {\bibfnamefont {T.}~\bibnamefont
  {Brandes}},\ }\bibfield  {title} {\enquote {\bibinfo {title} {Coherent and
  collective quantum optical effects in mesoscopic systems},}\ }\href
  {https://www.sciencedirect.com/science/article/abs/pii/S0370157304005496}
  {\bibfield  {journal} {\bibinfo  {journal} {Phys. Rep.}\ }\textbf {\bibinfo
  {volume} {408}},\ \bibinfo {pages} {315--474} (\bibinfo {year}
  {2005})}\BibitemShut {NoStop}%
\bibitem [{\citenamefont {Garraway}(2011)}]{Garraway2011}%
  \BibitemOpen
  \bibfield  {author} {\bibinfo {author} {\bibfnamefont {B.~M.}\ \bibnamefont
  {Garraway}},\ }\bibfield  {title} {\enquote {\bibinfo {title} {{The Dicke
  model in quantum optics: Dicke model revisited}},}\ }\href
  {https://royalsocietypublishing.org/doi/full/10.1098/rsta.2010.0333}
  {\bibfield  {journal} {\bibinfo  {journal} {Philos. Trans. R. Soc. A}\
  }\textbf {\bibinfo {volume} {369}},\ \bibinfo {pages} {1137--1155} (\bibinfo
  {year} {2011})}\BibitemShut {NoStop}%
\bibitem [{\citenamefont {Kirton}\ \emph {et~al.}(2019)\citenamefont {Kirton},
  \citenamefont {Roses}, \citenamefont {Keeling},\ and\ \citenamefont
  {Dalla~Torre}}]{Kirton2019}%
  \BibitemOpen
  \bibfield  {author} {\bibinfo {author} {\bibfnamefont {P.}~\bibnamefont
  {Kirton}}, \bibinfo {author} {\bibfnamefont {M.~M.}\ \bibnamefont {Roses}},
  \bibinfo {author} {\bibfnamefont {J.}~\bibnamefont {Keeling}}, \ and\
  \bibinfo {author} {\bibfnamefont {E.~G.}\ \bibnamefont {Dalla~Torre}},\
  }\bibfield  {title} {\enquote {\bibinfo {title} {{Introduction to the Dicke
  model: From equilibrium to nonequilibrium, and vice versa}},}\ }\href
  {https://onlinelibrary.wiley.com/doi/full/10.1002/qute.201800043?casa_token=fHpH24Oc00UAAAAA%3A3t_HL4M6vIyRF80_jpU8NBe1bdF6UliVLL1wRBFSgv61IXZ2uNku52ujWqAUmvaKPPJSG4f2NG8ZAu7QLQ}
  {\bibfield  {journal} {\bibinfo  {journal} {Adv. Quantum Technol.}\ }\textbf
  {\bibinfo {volume} {2}},\ \bibinfo {pages} {1800043} (\bibinfo {year}
  {2019})}\BibitemShut {NoStop}%
\bibitem [{\citenamefont {Fink}\ \emph {et~al.}(2009)\citenamefont {Fink},
  \citenamefont {Bianchetti}, \citenamefont {Baur}, \citenamefont {G\"oppl},
  \citenamefont {Steffen}, \citenamefont {Filipp}, \citenamefont {Leek},
  \citenamefont {Blais},\ and\ \citenamefont {Wallraff}}]{Fink2009}%
  \BibitemOpen
  \bibfield  {author} {\bibinfo {author} {\bibfnamefont {J.~M.}\ \bibnamefont
  {Fink}}, \bibinfo {author} {\bibfnamefont {R.}~\bibnamefont {Bianchetti}},
  \bibinfo {author} {\bibfnamefont {M.}~\bibnamefont {Baur}}, \bibinfo {author}
  {\bibfnamefont {M.}~\bibnamefont {G\"oppl}}, \bibinfo {author} {\bibfnamefont
  {L.}~\bibnamefont {Steffen}}, \bibinfo {author} {\bibfnamefont
  {S.}~\bibnamefont {Filipp}}, \bibinfo {author} {\bibfnamefont {P.~J.}\
  \bibnamefont {Leek}}, \bibinfo {author} {\bibfnamefont {A.}~\bibnamefont
  {Blais}}, \ and\ \bibinfo {author} {\bibfnamefont {A.}~\bibnamefont
  {Wallraff}},\ }\bibfield  {title} {\enquote {\bibinfo {title} {{Dressed
  collective qubit states and the Tavis-Cummings model in circuit QED}},}\
  }\href {\doibase 10.1103/PhysRevLett.103.083601} {\bibfield  {journal}
  {\bibinfo  {journal} {Phys. Rev. Lett.}\ }\textbf {\bibinfo {volume} {103}},\
  \bibinfo {pages} {083601} (\bibinfo {year} {2009})}\BibitemShut {NoStop}%
\bibitem [{\citenamefont {Feng}\ \emph {et~al.}(2015)\citenamefont {Feng},
  \citenamefont {Zhong}, \citenamefont {Liu}, \citenamefont {Yan},
  \citenamefont {Yang}, \citenamefont {Twamley},\ and\ \citenamefont
  {Wang}}]{Feng2015}%
  \BibitemOpen
  \bibfield  {author} {\bibinfo {author} {\bibfnamefont {M.}~\bibnamefont
  {Feng}}, \bibinfo {author} {\bibfnamefont {Y.~P.}\ \bibnamefont {Zhong}},
  \bibinfo {author} {\bibfnamefont {T.}~\bibnamefont {Liu}}, \bibinfo {author}
  {\bibfnamefont {L.~L.}\ \bibnamefont {Yan}}, \bibinfo {author} {\bibfnamefont
  {W.~L.}\ \bibnamefont {Yang}}, \bibinfo {author} {\bibfnamefont
  {J.}~\bibnamefont {Twamley}}, \ and\ \bibinfo {author} {\bibfnamefont
  {H.}~\bibnamefont {Wang}},\ }\bibfield  {title} {\enquote {\bibinfo {title}
  {{Exploring the quantum critical behaviour in a driven Tavis--Cummings
  circuit}},}\ }\href {https://www.nature.com/articles/ncomms8111} {\bibfield
  {journal} {\bibinfo  {journal} {Nat. Commun.}\ }\textbf {\bibinfo {volume}
  {6}},\ \bibinfo {pages} {1} (\bibinfo {year} {2015})}\BibitemShut {NoStop}%
\bibitem [{\citenamefont {De~Bernardis}\ \emph
  {et~al.}(2018{\natexlab{a}})\citenamefont {De~Bernardis}, \citenamefont
  {Jaako},\ and\ \citenamefont {Rabl}}]{DeBern2018}%
  \BibitemOpen
  \bibfield  {author} {\bibinfo {author} {\bibfnamefont {D.}~\bibnamefont
  {De~Bernardis}}, \bibinfo {author} {\bibfnamefont {T.}~\bibnamefont {Jaako}},
  \ and\ \bibinfo {author} {\bibfnamefont {P.}~\bibnamefont {Rabl}},\
  }\bibfield  {title} {\enquote {\bibinfo {title} {Cavity quantum
  electrodynamics in the nonperturbative regime},}\ }\href {\doibase
  10.1103/PhysRevA.97.043820} {\bibfield  {journal} {\bibinfo  {journal} {Phys.
  Rev. A}\ }\textbf {\bibinfo {volume} {97}},\ \bibinfo {pages} {043820}
  (\bibinfo {year} {2018}{\natexlab{a}})}\BibitemShut {NoStop}%
\bibitem [{\citenamefont {De~Bernardis}\ \emph
  {et~al.}(2018{\natexlab{b}})\citenamefont {De~Bernardis}, \citenamefont
  {Pilar}, \citenamefont {Jaako}, \citenamefont {De~Liberato},\ and\
  \citenamefont {Rabl}}]{DeBernardis2018}%
  \BibitemOpen
  \bibfield  {author} {\bibinfo {author} {\bibfnamefont {D.}~\bibnamefont
  {De~Bernardis}}, \bibinfo {author} {\bibfnamefont {P.}~\bibnamefont {Pilar}},
  \bibinfo {author} {\bibfnamefont {T.}~\bibnamefont {Jaako}}, \bibinfo
  {author} {\bibfnamefont {S.}~\bibnamefont {De~Liberato}}, \ and\ \bibinfo
  {author} {\bibfnamefont {P.}~\bibnamefont {Rabl}},\ }\bibfield  {title}
  {\enquote {\bibinfo {title} {Breakdown of gauge invariance in
  ultrastrong-coupling cavity {QED}},}\ }\href {\doibase
  10.1103/PhysRevA.98.053819} {\bibfield  {journal} {\bibinfo  {journal} {Phys.
  Rev. A}\ }\textbf {\bibinfo {volume} {98}},\ \bibinfo {pages} {053819}
  (\bibinfo {year} {2018}{\natexlab{b}})}\BibitemShut {NoStop}%
\bibitem [{\citenamefont {Stokes}\ and\ \citenamefont
  {Nazir}(2019)}]{Stokes2019}%
  \BibitemOpen
  \bibfield  {author} {\bibinfo {author} {\bibfnamefont {A.}~\bibnamefont
  {Stokes}}\ and\ \bibinfo {author} {\bibfnamefont {A.}~\bibnamefont {Nazir}},\
  }\bibfield  {title} {\enquote {\bibinfo {title} {Gauge ambiguities imply
  {J}aynes-{C}ummings physics remains valid in ultrastrong coupling {QED}},}\
  }\href {https://www.nature.com/articles/s41467-018-08101-0} {\bibfield
  {journal} {\bibinfo  {journal} {Nat. Commun.}\ }\textbf {\bibinfo {volume}
  {10}},\ \bibinfo {pages} {499} (\bibinfo {year} {2019})}\BibitemShut
  {NoStop}%
\bibitem [{\citenamefont {Kockum}\ \emph {et~al.}(2019)\citenamefont {Kockum},
  \citenamefont {Miranowicz}, \citenamefont {Liberato}, \citenamefont
  {Savasta},\ and\ \citenamefont {Nori}}]{Kockum2018}%
  \BibitemOpen
  \bibfield  {author} {\bibinfo {author} {\bibfnamefont {A.~F.}\ \bibnamefont
  {Kockum}}, \bibinfo {author} {\bibfnamefont {A.}~\bibnamefont {Miranowicz}},
  \bibinfo {author} {\bibfnamefont {S.~De}\ \bibnamefont {Liberato}}, \bibinfo
  {author} {\bibfnamefont {S.}~\bibnamefont {Savasta}}, \ and\ \bibinfo
  {author} {\bibfnamefont {F.}~\bibnamefont {Nori}},\ }\bibfield  {title}
  {\enquote {\bibinfo {title} {Ultrastrong coupling between light and
  matter},}\ }\href {\doibase 10.1038/s42254-018-0006-2} {\bibfield  {journal}
  {\bibinfo  {journal} {Nat. Rev. Phys.}\ }\textbf {\bibinfo {volume} {1}},\
  \bibinfo {pages} {19} (\bibinfo {year} {2019})}\BibitemShut {NoStop}%
\bibitem [{\citenamefont {Forn-D\'{\i}az}\ \emph {et~al.}(2019)\citenamefont
  {Forn-D\'{\i}az}, \citenamefont {Lamata}, \citenamefont {Rico}, \citenamefont
  {Kono},\ and\ \citenamefont {Solano}}]{Forn-Diaz2018}%
  \BibitemOpen
  \bibfield  {author} {\bibinfo {author} {\bibfnamefont {P.}~\bibnamefont
  {Forn-D\'{\i}az}}, \bibinfo {author} {\bibfnamefont {L.}~\bibnamefont
  {Lamata}}, \bibinfo {author} {\bibfnamefont {E.}~\bibnamefont {Rico}},
  \bibinfo {author} {\bibfnamefont {J.}~\bibnamefont {Kono}}, \ and\ \bibinfo
  {author} {\bibfnamefont {E.}~\bibnamefont {Solano}},\ }\bibfield  {title}
  {\enquote {\bibinfo {title} {Ultrastrong coupling regimes of light-matter
  interaction},}\ }\href {\doibase 10.1103/RevModPhys.91.025005} {\bibfield
  {journal} {\bibinfo  {journal} {Rev. Mod. Phys.}\ }\textbf {\bibinfo {volume}
  {91}},\ \bibinfo {pages} {025005} (\bibinfo {year} {2019})}\BibitemShut
  {NoStop}%
\bibitem [{\citenamefont {Schwartz}\ \emph {et~al.}(2011)\citenamefont
  {Schwartz}, \citenamefont {Hutchison}, \citenamefont {Genet},\ and\
  \citenamefont {Ebbesen}}]{Schwartz2011}%
  \BibitemOpen
  \bibfield  {author} {\bibinfo {author} {\bibfnamefont {T.}~\bibnamefont
  {Schwartz}}, \bibinfo {author} {\bibfnamefont {J.~A.}\ \bibnamefont
  {Hutchison}}, \bibinfo {author} {\bibfnamefont {C.}~\bibnamefont {Genet}}, \
  and\ \bibinfo {author} {\bibfnamefont {T.~W.}\ \bibnamefont {Ebbesen}},\
  }\bibfield  {title} {\enquote {\bibinfo {title} {Reversible switching of
  ultrastrong light-molecule coupling},}\ }\href {\doibase
  10.1103/PhysRevLett.106.196405} {\bibfield  {journal} {\bibinfo  {journal}
  {Phys. Rev. Lett.}\ }\textbf {\bibinfo {volume} {106}},\ \bibinfo {pages}
  {196405} (\bibinfo {year} {2011})}\BibitemShut {NoStop}%
\bibitem [{\citenamefont {K{\'{e}}na-Cohen}\ \emph {et~al.}(2013)\citenamefont
  {K{\'{e}}na-Cohen}, \citenamefont {Maier},\ and\ \citenamefont
  {Bradley}}]{Kena-Cohen2013}%
  \BibitemOpen
  \bibfield  {author} {\bibinfo {author} {\bibfnamefont {S.}~\bibnamefont
  {K{\'{e}}na-Cohen}}, \bibinfo {author} {\bibfnamefont {S.~A.}\ \bibnamefont
  {Maier}}, \ and\ \bibinfo {author} {\bibfnamefont {D.~D.~C.}\ \bibnamefont
  {Bradley}},\ }\bibfield  {title} {\enquote {\bibinfo {title} {Ultrastrongly
  coupled exciton-polaritons in metal-clad organic semiconductor
  microcavities},}\ }\href {\doibase 10.1002/adom.201300256} {\bibfield
  {journal} {\bibinfo  {journal} {Adv. Opt. Mater.}\ }\textbf {\bibinfo
  {volume} {1}},\ \bibinfo {pages} {827} (\bibinfo {year} {2013})}\BibitemShut
  {NoStop}%
\bibitem [{\citenamefont {Gubbin}\ \emph {et~al.}(2014)\citenamefont {Gubbin},
  \citenamefont {Maier},\ and\ \citenamefont {K{\'{e}}na-Cohen}}]{Gubbin2014}%
  \BibitemOpen
  \bibfield  {author} {\bibinfo {author} {\bibfnamefont {C.~R.}\ \bibnamefont
  {Gubbin}}, \bibinfo {author} {\bibfnamefont {S.~A.}\ \bibnamefont {Maier}}, \
  and\ \bibinfo {author} {\bibfnamefont {S.}~\bibnamefont {K{\'{e}}na-Cohen}},\
  }\bibfield  {title} {\enquote {\bibinfo {title} {{Low-voltage polariton
  electroluminescence from an ultrastrongly coupled organic light-emitting
  diode}},}\ }\href {\doibase 10.1063/1.4871271} {\bibfield  {journal}
  {\bibinfo  {journal} {Appl. Phys. Lett.}\ }\textbf {\bibinfo {volume}
  {104}},\ \bibinfo {pages} {233302} (\bibinfo {year} {2014})}\BibitemShut
  {NoStop}%
\bibitem [{\citenamefont {Mazzeo}\ \emph {et~al.}(2014)\citenamefont {Mazzeo},
  \citenamefont {Genco}, \citenamefont {Gambino}, \citenamefont {Ballarini},
  \citenamefont {Mangione}, \citenamefont {{Di Stefano}}, \citenamefont
  {Patan{\`{e}}}, \citenamefont {Savasta}, \citenamefont {Sanvitto},\ and\
  \citenamefont {Gigli}}]{Mazzeo2014}%
  \BibitemOpen
  \bibfield  {author} {\bibinfo {author} {\bibfnamefont {M.}~\bibnamefont
  {Mazzeo}}, \bibinfo {author} {\bibfnamefont {A.}~\bibnamefont {Genco}},
  \bibinfo {author} {\bibfnamefont {S.}~\bibnamefont {Gambino}}, \bibinfo
  {author} {\bibfnamefont {D.}~\bibnamefont {Ballarini}}, \bibinfo {author}
  {\bibfnamefont {F.}~\bibnamefont {Mangione}}, \bibinfo {author}
  {\bibfnamefont {O.}~\bibnamefont {{Di Stefano}}}, \bibinfo {author}
  {\bibfnamefont {S.}~\bibnamefont {Patan{\`{e}}}}, \bibinfo {author}
  {\bibfnamefont {S.}~\bibnamefont {Savasta}}, \bibinfo {author} {\bibfnamefont
  {D.}~\bibnamefont {Sanvitto}}, \ and\ \bibinfo {author} {\bibfnamefont
  {G.}~\bibnamefont {Gigli}},\ }\bibfield  {title} {\enquote {\bibinfo {title}
  {{Ultrastrong light-matter coupling in electrically doped microcavity organic
  light emitting diodes}},}\ }\href
  {http://aip.scitation.org/doi/10.1063/1.4882422} {\bibfield  {journal}
  {\bibinfo  {journal} {Appl. Phys. Lett.}\ }\textbf {\bibinfo {volume}
  {104}},\ \bibinfo {pages} {233303} (\bibinfo {year} {2014})}\BibitemShut
  {NoStop}%
\bibitem [{\citenamefont {Gambino}\ \emph {et~al.}(2014)\citenamefont
  {Gambino}, \citenamefont {Mazzeo}, \citenamefont {Genco}, \citenamefont {{Di
  Stefano}}, \citenamefont {Savasta}, \citenamefont {Patan{\`{e}}},
  \citenamefont {Ballarini}, \citenamefont {Mangione}, \citenamefont {Lerario},
  \citenamefont {Sanvitto},\ and\ \citenamefont {Gigli}}]{Gambino2014}%
  \BibitemOpen
  \bibfield  {author} {\bibinfo {author} {\bibfnamefont {S.}~\bibnamefont
  {Gambino}}, \bibinfo {author} {\bibfnamefont {M.}~\bibnamefont {Mazzeo}},
  \bibinfo {author} {\bibfnamefont {A.}~\bibnamefont {Genco}}, \bibinfo
  {author} {\bibfnamefont {O.}~\bibnamefont {{Di Stefano}}}, \bibinfo {author}
  {\bibfnamefont {S.}~\bibnamefont {Savasta}}, \bibinfo {author} {\bibfnamefont
  {S.}~\bibnamefont {Patan{\`{e}}}}, \bibinfo {author} {\bibfnamefont
  {D.}~\bibnamefont {Ballarini}}, \bibinfo {author} {\bibfnamefont
  {F.}~\bibnamefont {Mangione}}, \bibinfo {author} {\bibfnamefont
  {G.}~\bibnamefont {Lerario}}, \bibinfo {author} {\bibfnamefont
  {D.}~\bibnamefont {Sanvitto}}, \ and\ \bibinfo {author} {\bibfnamefont
  {G.}~\bibnamefont {Gigli}},\ }\bibfield  {title} {\enquote {\bibinfo {title}
  {Exploring light--matter interaction phenomena under ultrastrong coupling
  regime},}\ }\href {\doibase 10.1021/ph500266d} {\bibfield  {journal}
  {\bibinfo  {journal} {ACS Photonics}\ }\textbf {\bibinfo {volume} {1}},\
  \bibinfo {pages} {1042} (\bibinfo {year} {2014})}\BibitemShut {NoStop}%
\bibitem [{\citenamefont {Genco}\ \emph {et~al.}(2018)\citenamefont {Genco},
  \citenamefont {Ridolfo}, \citenamefont {Savasta}, \citenamefont {Patan{\`e}},
  \citenamefont {Gigli},\ and\ \citenamefont {Mazzeo}}]{Genco2018}%
  \BibitemOpen
  \bibfield  {author} {\bibinfo {author} {\bibfnamefont {A.}~\bibnamefont
  {Genco}}, \bibinfo {author} {\bibfnamefont {A.}~\bibnamefont {Ridolfo}},
  \bibinfo {author} {\bibfnamefont {S.}~\bibnamefont {Savasta}}, \bibinfo
  {author} {\bibfnamefont {S.}~\bibnamefont {Patan{\`e}}}, \bibinfo {author}
  {\bibfnamefont {G.}~\bibnamefont {Gigli}}, \ and\ \bibinfo {author}
  {\bibfnamefont {M.}~\bibnamefont {Mazzeo}},\ }\bibfield  {title} {\enquote
  {\bibinfo {title} {{Bright polariton Coumarin-based OLEDs operating in the
  ultrastrong coupling regime}},}\ }\href
  {https://onlinelibrary.wiley.com/doi/abs/10.1002/adom.201800364} {\bibfield
  {journal} {\bibinfo  {journal} {Adv. Opt. Mater.}\ }\textbf {\bibinfo
  {volume} {6}},\ \bibinfo {pages} {1800364} (\bibinfo {year}
  {2018})}\BibitemShut {NoStop}%
\bibitem [{\citenamefont {Anappara}\ \emph {et~al.}(2009)\citenamefont
  {Anappara}, \citenamefont {{De Liberato}}, \citenamefont {Tredicucci},
  \citenamefont {Ciuti}, \citenamefont {Biasiol}, \citenamefont {Sorba},\ and\
  \citenamefont {Beltram}}]{Anappara2009}%
  \BibitemOpen
  \bibfield  {author} {\bibinfo {author} {\bibfnamefont {A.~A.}\ \bibnamefont
  {Anappara}}, \bibinfo {author} {\bibfnamefont {S.}~\bibnamefont {{De
  Liberato}}}, \bibinfo {author} {\bibfnamefont {A.}~\bibnamefont
  {Tredicucci}}, \bibinfo {author} {\bibfnamefont {C.}~\bibnamefont {Ciuti}},
  \bibinfo {author} {\bibfnamefont {G.}~\bibnamefont {Biasiol}}, \bibinfo
  {author} {\bibfnamefont {L.}~\bibnamefont {Sorba}}, \ and\ \bibinfo {author}
  {\bibfnamefont {F.}~\bibnamefont {Beltram}},\ }\bibfield  {title} {\enquote
  {\bibinfo {title} {{Signatures of the ultrastrong light-matter coupling
  regime}},}\ }\href {http://link.aps.org/doi/10.1103/PhysRevB.79.201303}
  {\bibfield  {journal} {\bibinfo  {journal} {Phys. Rev. B}\ }\textbf {\bibinfo
  {volume} {79}},\ \bibinfo {pages} {201303} (\bibinfo {year}
  {2009})}\BibitemShut {NoStop}%
\bibitem [{\citenamefont {G{\"{u}}nter}\ \emph {et~al.}(2009)\citenamefont
  {G{\"{u}}nter}, \citenamefont {Anappara}, \citenamefont {Hees}, \citenamefont
  {Sell}, \citenamefont {Biasiol}, \citenamefont {Sorba}, \citenamefont {{De
  Liberato}}, \citenamefont {Ciuti}, \citenamefont {Tredicucci}, \citenamefont
  {Leitenstorfer},\ and\ \citenamefont {Huber}}]{Gunter2009}%
  \BibitemOpen
  \bibfield  {author} {\bibinfo {author} {\bibfnamefont {G.}~\bibnamefont
  {G{\"{u}}nter}}, \bibinfo {author} {\bibfnamefont {A.~A.}\ \bibnamefont
  {Anappara}}, \bibinfo {author} {\bibfnamefont {J.}~\bibnamefont {Hees}},
  \bibinfo {author} {\bibfnamefont {A.}~\bibnamefont {Sell}}, \bibinfo {author}
  {\bibfnamefont {G.}~\bibnamefont {Biasiol}}, \bibinfo {author} {\bibfnamefont
  {L.}~\bibnamefont {Sorba}}, \bibinfo {author} {\bibfnamefont
  {S.}~\bibnamefont {{De Liberato}}}, \bibinfo {author} {\bibfnamefont
  {C.}~\bibnamefont {Ciuti}}, \bibinfo {author} {\bibfnamefont
  {A.}~\bibnamefont {Tredicucci}}, \bibinfo {author} {\bibfnamefont
  {A.}~\bibnamefont {Leitenstorfer}}, \ and\ \bibinfo {author} {\bibfnamefont
  {R.}~\bibnamefont {Huber}},\ }\bibfield  {title} {\enquote {\bibinfo {title}
  {{Sub-cycle switch-on of ultrastrong light--matter interaction}},}\ }\href
  {\doibase 10.1038/nature07838} {\bibfield  {journal} {\bibinfo  {journal}
  {Nature}\ }\textbf {\bibinfo {volume} {458}},\ \bibinfo {pages} {178}
  (\bibinfo {year} {2009})}\BibitemShut {NoStop}%
\bibitem [{\citenamefont {Todorov}\ \emph {et~al.}(2010)\citenamefont
  {Todorov}, \citenamefont {Andrews}, \citenamefont {Colombelli}, \citenamefont
  {{De Liberato}}, \citenamefont {Ciuti}, \citenamefont {Klang}, \citenamefont
  {Strasser},\ and\ \citenamefont {Sirtori}}]{Todorov2010}%
  \BibitemOpen
  \bibfield  {author} {\bibinfo {author} {\bibfnamefont {Y.}~\bibnamefont
  {Todorov}}, \bibinfo {author} {\bibfnamefont {A.~M.}\ \bibnamefont
  {Andrews}}, \bibinfo {author} {\bibfnamefont {R.}~\bibnamefont {Colombelli}},
  \bibinfo {author} {\bibfnamefont {S.}~\bibnamefont {{De Liberato}}}, \bibinfo
  {author} {\bibfnamefont {C.}~\bibnamefont {Ciuti}}, \bibinfo {author}
  {\bibfnamefont {P.}~\bibnamefont {Klang}}, \bibinfo {author} {\bibfnamefont
  {G.}~\bibnamefont {Strasser}}, \ and\ \bibinfo {author} {\bibfnamefont
  {C.}~\bibnamefont {Sirtori}},\ }\bibfield  {title} {\enquote {\bibinfo
  {title} {Ultrastrong light-matter coupling regime with polariton dots},}\
  }\href {http://link.aps.org/doi/10.1103/PhysRevLett.105.196402} {\bibfield
  {journal} {\bibinfo  {journal} {Phys. Rev. Lett.}\ }\textbf {\bibinfo
  {volume} {105}},\ \bibinfo {pages} {196402} (\bibinfo {year}
  {2010})}\BibitemShut {NoStop}%
\bibitem [{\citenamefont {Askenazi}\ \emph {et~al.}(2014)\citenamefont
  {Askenazi}, \citenamefont {Vasanelli}, \citenamefont {Delteil}, \citenamefont
  {Todorov}, \citenamefont {Andreani}, \citenamefont {Beaudoin}, \citenamefont
  {Sagnes},\ and\ \citenamefont {Sirtori}}]{Askenazi2014}%
  \BibitemOpen
  \bibfield  {author} {\bibinfo {author} {\bibfnamefont {B.}~\bibnamefont
  {Askenazi}}, \bibinfo {author} {\bibfnamefont {A.}~\bibnamefont {Vasanelli}},
  \bibinfo {author} {\bibfnamefont {A.}~\bibnamefont {Delteil}}, \bibinfo
  {author} {\bibfnamefont {Y.}~\bibnamefont {Todorov}}, \bibinfo {author}
  {\bibfnamefont {L.~C.}\ \bibnamefont {Andreani}}, \bibinfo {author}
  {\bibfnamefont {G.}~\bibnamefont {Beaudoin}}, \bibinfo {author}
  {\bibfnamefont {I.}~\bibnamefont {Sagnes}}, \ and\ \bibinfo {author}
  {\bibfnamefont {C.}~\bibnamefont {Sirtori}},\ }\bibfield  {title} {\enquote
  {\bibinfo {title} {{Ultra-strong light-matter coupling for designer
  Reststrahlen band}},}\ }\href {\doibase 10.1088/1367-2630/16/4/043029}
  {\bibfield  {journal} {\bibinfo  {journal} {New J. Phys.}\ }\textbf {\bibinfo
  {volume} {16}},\ \bibinfo {pages} {043029} (\bibinfo {year}
  {2014})}\BibitemShut {NoStop}%
\bibitem [{\citenamefont {Scalari}\ \emph {et~al.}(2012)\citenamefont
  {Scalari}, \citenamefont {Maissen}, \citenamefont {Turcinkova}, \citenamefont
  {Hagenmuller}, \citenamefont {{De Liberato}}, \citenamefont {Ciuti},
  \citenamefont {Reichl}, \citenamefont {Schuh}, \citenamefont {Wegscheider},
  \citenamefont {Beck},\ and\ \citenamefont {Faist}}]{Scalari2012}%
  \BibitemOpen
  \bibfield  {author} {\bibinfo {author} {\bibfnamefont {G.}~\bibnamefont
  {Scalari}}, \bibinfo {author} {\bibfnamefont {C.}~\bibnamefont {Maissen}},
  \bibinfo {author} {\bibfnamefont {D.}~\bibnamefont {Turcinkova}}, \bibinfo
  {author} {\bibfnamefont {D.}~\bibnamefont {Hagenmuller}}, \bibinfo {author}
  {\bibfnamefont {S.}~\bibnamefont {{De Liberato}}}, \bibinfo {author}
  {\bibfnamefont {C.}~\bibnamefont {Ciuti}}, \bibinfo {author} {\bibfnamefont
  {C.}~\bibnamefont {Reichl}}, \bibinfo {author} {\bibfnamefont
  {D.}~\bibnamefont {Schuh}}, \bibinfo {author} {\bibfnamefont
  {W.}~\bibnamefont {Wegscheider}}, \bibinfo {author} {\bibfnamefont
  {M.}~\bibnamefont {Beck}}, \ and\ \bibinfo {author} {\bibfnamefont
  {J.}~\bibnamefont {Faist}},\ }\bibfield  {title} {\enquote {\bibinfo {title}
  {{Ultrastrong coupling of the cyclotron transition of a 2D electron gas to a
  THz metamaterial}},}\ }\href
  {http://www.sciencemag.org/cgi/doi/10.1126/science.1216022} {\bibfield
  {journal} {\bibinfo  {journal} {Science}\ }\textbf {\bibinfo {volume}
  {335}},\ \bibinfo {pages} {1323} (\bibinfo {year} {2012})}\BibitemShut
  {NoStop}%
\bibitem [{\citenamefont {Maissen}\ \emph {et~al.}(2014)\citenamefont
  {Maissen}, \citenamefont {Scalari}, \citenamefont {Valmorra}, \citenamefont
  {Beck}, \citenamefont {Faist}, \citenamefont {Cibella}, \citenamefont
  {Leoni}, \citenamefont {Reichl}, \citenamefont {Charpentier},\ and\
  \citenamefont {Wegscheider}}]{Maissen2014}%
  \BibitemOpen
  \bibfield  {author} {\bibinfo {author} {\bibfnamefont {C.}~\bibnamefont
  {Maissen}}, \bibinfo {author} {\bibfnamefont {G.}~\bibnamefont {Scalari}},
  \bibinfo {author} {\bibfnamefont {F.}~\bibnamefont {Valmorra}}, \bibinfo
  {author} {\bibfnamefont {M.}~\bibnamefont {Beck}}, \bibinfo {author}
  {\bibfnamefont {J.}~\bibnamefont {Faist}}, \bibinfo {author} {\bibfnamefont
  {S.}~\bibnamefont {Cibella}}, \bibinfo {author} {\bibfnamefont
  {R.}~\bibnamefont {Leoni}}, \bibinfo {author} {\bibfnamefont
  {C.}~\bibnamefont {Reichl}}, \bibinfo {author} {\bibfnamefont
  {C.}~\bibnamefont {Charpentier}}, \ and\ \bibinfo {author} {\bibfnamefont
  {W.}~\bibnamefont {Wegscheider}},\ }\bibfield  {title} {\enquote {\bibinfo
  {title} {{Ultrastrong coupling in the near field of complementary split-ring
  resonators}},}\ }\href {http://dx.doi.org/10.1103/PhysRevB.90.205309}
  {\bibfield  {journal} {\bibinfo  {journal} {Phys. Rev. B}\ }\textbf {\bibinfo
  {volume} {90}},\ \bibinfo {pages} {205309} (\bibinfo {year}
  {2014})}\BibitemShut {NoStop}%
\bibitem [{\citenamefont {Zhang}\ \emph {et~al.}(2016)\citenamefont {Zhang},
  \citenamefont {Lou}, \citenamefont {Li}, \citenamefont {Reno}, \citenamefont
  {Pan}, \citenamefont {Watson}, \citenamefont {Manfra},\ and\ \citenamefont
  {Kono}}]{Zhang2016a}%
  \BibitemOpen
  \bibfield  {author} {\bibinfo {author} {\bibfnamefont {Q.}~\bibnamefont
  {Zhang}}, \bibinfo {author} {\bibfnamefont {M.}~\bibnamefont {Lou}}, \bibinfo
  {author} {\bibfnamefont {X.}~\bibnamefont {Li}}, \bibinfo {author}
  {\bibfnamefont {J.~L.}\ \bibnamefont {Reno}}, \bibinfo {author}
  {\bibfnamefont {W.}~\bibnamefont {Pan}}, \bibinfo {author} {\bibfnamefont
  {J.~D.}\ \bibnamefont {Watson}}, \bibinfo {author} {\bibfnamefont {M.~J.}\
  \bibnamefont {Manfra}}, \ and\ \bibinfo {author} {\bibfnamefont
  {J.}~\bibnamefont {Kono}},\ }\bibfield  {title} {\enquote {\bibinfo {title}
  {{Collective non-perturbative coupling of 2D electrons with
  high-quality-factor terahertz cavity photons}},}\ }\href
  {http://dx.doi.org/10.1038/nphys3850} {\bibfield  {journal} {\bibinfo
  {journal} {Nat. Phys.}\ }\textbf {\bibinfo {volume} {12}},\ \bibinfo {pages}
  {1005} (\bibinfo {year} {2016})}\BibitemShut {NoStop}%
\bibitem [{\citenamefont {Bayer}\ \emph {et~al.}(2017)\citenamefont {Bayer},
  \citenamefont {Pozimski}, \citenamefont {Schambeck}, \citenamefont {Schuh},
  \citenamefont {Huber}, \citenamefont {Bougeard},\ and\ \citenamefont
  {Lange}}]{Bayer2017}%
  \BibitemOpen
  \bibfield  {author} {\bibinfo {author} {\bibfnamefont {A.}~\bibnamefont
  {Bayer}}, \bibinfo {author} {\bibfnamefont {M.}~\bibnamefont {Pozimski}},
  \bibinfo {author} {\bibfnamefont {S.}~\bibnamefont {Schambeck}}, \bibinfo
  {author} {\bibfnamefont {D.}~\bibnamefont {Schuh}}, \bibinfo {author}
  {\bibfnamefont {R.}~\bibnamefont {Huber}}, \bibinfo {author} {\bibfnamefont
  {D.}~\bibnamefont {Bougeard}}, \ and\ \bibinfo {author} {\bibfnamefont
  {C.}~\bibnamefont {Lange}},\ }\bibfield  {title} {\enquote {\bibinfo {title}
  {Terahertz light-matter interaction beyond unity coupling strength},}\ }\href
  {\doibase 10.1021/acs.nanolett.7b03103} {\bibfield  {journal} {\bibinfo
  {journal} {Nano Lett.}\ }\textbf {\bibinfo {volume} {17}},\ \bibinfo {pages}
  {6340} (\bibinfo {year} {2017})}\BibitemShut {NoStop}%
\bibitem [{\citenamefont {Li}\ \emph {et~al.}(2018)\citenamefont {Li},
  \citenamefont {Bamba}, \citenamefont {Zhang}, \citenamefont {Fallahi},
  \citenamefont {Gardner}, \citenamefont {Gao}, \citenamefont {Lou},
  \citenamefont {Yoshioka}, \citenamefont {Manfra},\ and\ \citenamefont
  {Kono}}]{Li2018}%
  \BibitemOpen
  \bibfield  {author} {\bibinfo {author} {\bibfnamefont {X.}~\bibnamefont
  {Li}}, \bibinfo {author} {\bibfnamefont {M.}~\bibnamefont {Bamba}}, \bibinfo
  {author} {\bibfnamefont {Q.}~\bibnamefont {Zhang}}, \bibinfo {author}
  {\bibfnamefont {S.}~\bibnamefont {Fallahi}}, \bibinfo {author} {\bibfnamefont
  {G.~C.}\ \bibnamefont {Gardner}}, \bibinfo {author} {\bibfnamefont
  {W.}~\bibnamefont {Gao}}, \bibinfo {author} {\bibfnamefont {M.}~\bibnamefont
  {Lou}}, \bibinfo {author} {\bibfnamefont {K.}~\bibnamefont {Yoshioka}},
  \bibinfo {author} {\bibfnamefont {M.~J.}\ \bibnamefont {Manfra}}, \ and\
  \bibinfo {author} {\bibfnamefont {J.}~\bibnamefont {Kono}},\ }\bibfield
  {title} {\enquote {\bibinfo {title} {{Vacuum Bloch--Siegert shift in Landau
  polaritons with ultra-high cooperativity}},}\ }\href
  {https://www.nature.com/articles/s41566-018-0153-0} {\bibfield  {journal}
  {\bibinfo  {journal} {Nat. Photonics}\ }\textbf {\bibinfo {volume} {12}},\
  \bibinfo {pages} {324--329} (\bibinfo {year} {2018})}\BibitemShut {NoStop}%
\bibitem [{\citenamefont {Di~Stefano}\ \emph {et~al.}(2019)\citenamefont
  {Di~Stefano}, \citenamefont {Settineri}, \citenamefont {Macr{\`\i}},
  \citenamefont {Garziano}, \citenamefont {Stassi}, \citenamefont {Savasta},\
  and\ \citenamefont {Nori}}]{DiStefano2019}%
  \BibitemOpen
  \bibfield  {author} {\bibinfo {author} {\bibfnamefont {O.}~\bibnamefont
  {Di~Stefano}}, \bibinfo {author} {\bibfnamefont {A.}~\bibnamefont
  {Settineri}}, \bibinfo {author} {\bibfnamefont {V.}~\bibnamefont
  {Macr{\`\i}}}, \bibinfo {author} {\bibfnamefont {L.}~\bibnamefont
  {Garziano}}, \bibinfo {author} {\bibfnamefont {R.}~\bibnamefont {Stassi}},
  \bibinfo {author} {\bibfnamefont {S.}~\bibnamefont {Savasta}}, \ and\
  \bibinfo {author} {\bibfnamefont {F.}~\bibnamefont {Nori}},\ }\bibfield
  {title} {\enquote {\bibinfo {title} {{Resolution of gauge ambiguities in
  ultrastrong-coupling cavity QED}},}\ }\href
  {https://arxiv.org/abs/1809.08749} {\bibfield  {journal} {\bibinfo  {journal}
  {Nat. Phys.}\ }\textbf {\bibinfo {volume} {15}},\ \bibinfo {pages} {803}
  (\bibinfo {year} {2019})}\BibitemShut {NoStop}%
\bibitem [{\citenamefont {Starace}(1971)}]{Starace1971}%
  \BibitemOpen
  \bibfield  {author} {\bibinfo {author} {\bibfnamefont {A.~F.}\ \bibnamefont
  {Starace}},\ }\bibfield  {title} {\enquote {\bibinfo {title} {Length and
  velocity formulas in approximate oscillator-strength calculations},}\ }\href
  {\doibase 10.1103/PhysRevA.3.1242} {\bibfield  {journal} {\bibinfo  {journal}
  {Phys. Rev. A}\ }\textbf {\bibinfo {volume} {3}},\ \bibinfo {pages}
  {1242--1245} (\bibinfo {year} {1971})}\BibitemShut {NoStop}%
\bibitem [{\citenamefont {Girlanda}\ \emph {et~al.}(1981)\citenamefont
  {Girlanda}, \citenamefont {Quattropani},\ and\ \citenamefont
  {Schwendimann}}]{Girlanda1981}%
  \BibitemOpen
  \bibfield  {author} {\bibinfo {author} {\bibfnamefont {R.}~\bibnamefont
  {Girlanda}}, \bibinfo {author} {\bibfnamefont {A.}~\bibnamefont
  {Quattropani}}, \ and\ \bibinfo {author} {\bibfnamefont {P.}~\bibnamefont
  {Schwendimann}},\ }\bibfield  {title} {\enquote {\bibinfo {title} {Two-photon
  transitions to exciton states in semiconductors. application to {C}u{C}l},}\
  }\href {\doibase 10.1103/PhysRevB.24.2009} {\bibfield  {journal} {\bibinfo
  {journal} {Phys. Rev. B}\ }\textbf {\bibinfo {volume} {24}},\ \bibinfo
  {pages} {2009} (\bibinfo {year} {1981})}\BibitemShut {NoStop}%
\bibitem [{\citenamefont {Ismail-Beigi}\ \emph {et~al.}(2001)\citenamefont
  {Ismail-Beigi}, \citenamefont {Chang},\ and\ \citenamefont
  {Louie}}]{Ismail-Beigi2001}%
  \BibitemOpen
  \bibfield  {author} {\bibinfo {author} {\bibfnamefont {S.}~\bibnamefont
  {Ismail-Beigi}}, \bibinfo {author} {\bibfnamefont {E.~K.}\ \bibnamefont
  {Chang}}, \ and\ \bibinfo {author} {\bibfnamefont {S.~G.}\ \bibnamefont
  {Louie}},\ }\bibfield  {title} {\enquote {\bibinfo {title} {Coupling of
  nonlocal potentials to electromagnetic fields},}\ }\href {\doibase
  10.1103/PhysRevLett.87.087402} {\bibfield  {journal} {\bibinfo  {journal}
  {Phys. Rev. Lett.}\ }\textbf {\bibinfo {volume} {87}},\ \bibinfo {pages}
  {087402} (\bibinfo {year} {2001})}\BibitemShut {NoStop}%
\bibitem [{\citenamefont {Settineri}\ \emph {et~al.}(2019)\citenamefont
  {Settineri}, \citenamefont {Di~Stefano}, \citenamefont {Zueco}, \citenamefont
  {Hughes}, \citenamefont {Savasta},\ and\ \citenamefont
  {Nori}}]{Settineri2020}%
  \BibitemOpen
  \bibfield  {author} {\bibinfo {author} {\bibfnamefont {A.}~\bibnamefont
  {Settineri}}, \bibinfo {author} {\bibfnamefont {O.}~\bibnamefont
  {Di~Stefano}}, \bibinfo {author} {\bibfnamefont {D.}~\bibnamefont {Zueco}},
  \bibinfo {author} {\bibfnamefont {S.}~\bibnamefont {Hughes}}, \bibinfo
  {author} {\bibfnamefont {S.}~\bibnamefont {Savasta}}, \ and\ \bibinfo
  {author} {\bibfnamefont {F.}~\bibnamefont {Nori}},\ }\bibfield  {title}
  {\enquote {\bibinfo {title} {Gauge freedom, quantum measurements, and
  time-dependent interactions in cavity and circuit {QED}},}\ }\href
  {https://arxiv.org/abs/1912.08548} {\bibfield  {journal} {\bibinfo  {journal}
  {arxiv: 1912.08548}\ } (\bibinfo {year} {2019})}\BibitemShut {NoStop}%
\bibitem [{\citenamefont {Emary}\ and\ \citenamefont
  {Brandes}(2003)}]{Emary2003}%
  \BibitemOpen
  \bibfield  {author} {\bibinfo {author} {\bibfnamefont {C.}~\bibnamefont
  {Emary}}\ and\ \bibinfo {author} {\bibfnamefont {T.}~\bibnamefont
  {Brandes}},\ }\bibfield  {title} {\enquote {\bibinfo {title} {{Quantum chaos
  triggered by precursors of a quantum phase transition: The Dicke model}},}\
  }\href {\doibase 10.1103/PhysRevLett.90.044101} {\bibfield  {journal}
  {\bibinfo  {journal} {Phys. Rev. Lett.}\ }\textbf {\bibinfo {volume} {90}},\
  \bibinfo {pages} {044101} (\bibinfo {year} {2003})}\BibitemShut {NoStop}%
\bibitem [{\citenamefont {Lambert}\ \emph {et~al.}(2004)\citenamefont
  {Lambert}, \citenamefont {Emary},\ and\ \citenamefont
  {Brandes}}]{Lambert2004}%
  \BibitemOpen
  \bibfield  {author} {\bibinfo {author} {\bibfnamefont {N.}~\bibnamefont
  {Lambert}}, \bibinfo {author} {\bibfnamefont {C.}~\bibnamefont {Emary}}, \
  and\ \bibinfo {author} {\bibfnamefont {T.}~\bibnamefont {Brandes}},\
  }\bibfield  {title} {\enquote {\bibinfo {title} {Entanglement and the phase
  transition in single-mode superradiance},}\ }\href {\doibase
  10.1103/PhysRevLett.92.073602} {\bibfield  {journal} {\bibinfo  {journal}
  {Phys. Rev. Lett.}\ }\textbf {\bibinfo {volume} {92}},\ \bibinfo {pages}
  {073602} (\bibinfo {year} {2004})}\BibitemShut {NoStop}%
\bibitem [{\citenamefont {Shammah}\ \emph {et~al.}(2017)\citenamefont
  {Shammah}, \citenamefont {Lambert}, \citenamefont {Nori},\ and\ \citenamefont
  {De~Liberato}}]{Nathan2017}%
  \BibitemOpen
  \bibfield  {author} {\bibinfo {author} {\bibfnamefont {N.}~\bibnamefont
  {Shammah}}, \bibinfo {author} {\bibfnamefont {N.}~\bibnamefont {Lambert}},
  \bibinfo {author} {\bibfnamefont {F.}~\bibnamefont {Nori}}, \ and\ \bibinfo
  {author} {\bibfnamefont {S.}~\bibnamefont {De~Liberato}},\ }\bibfield
  {title} {\enquote {\bibinfo {title} {Superradiance with local phase-breaking
  effects},}\ }\href@noop {} {\bibfield  {journal} {\bibinfo  {journal} {Phys.
  Rev. A}\ }\textbf {\bibinfo {volume} {96}},\ \bibinfo {pages} {023863}
  (\bibinfo {year} {2017})}\BibitemShut {NoStop}%
\bibitem [{\citenamefont {Shammah}\ \emph {et~al.}(2018)\citenamefont
  {Shammah}, \citenamefont {Ahmed}, \citenamefont {Lambert}, \citenamefont
  {De~Liberato},\ and\ \citenamefont {Nori}}]{Nathan2018}%
  \BibitemOpen
  \bibfield  {author} {\bibinfo {author} {\bibfnamefont {N.}~\bibnamefont
  {Shammah}}, \bibinfo {author} {\bibfnamefont {S.}~\bibnamefont {Ahmed}},
  \bibinfo {author} {\bibfnamefont {N.}~\bibnamefont {Lambert}}, \bibinfo
  {author} {\bibfnamefont {S.}~\bibnamefont {De~Liberato}}, \ and\ \bibinfo
  {author} {\bibfnamefont {F.}~\bibnamefont {Nori}},\ }\bibfield  {title}
  {\enquote {\bibinfo {title} {Open quantum systems with local and collective
  incoherent processes: Efficient numerical simulations using permutational
  invariance},}\ }\href@noop {} {\bibfield  {journal} {\bibinfo  {journal}
  {Phys. Rev. A}\ }\textbf {\bibinfo {volume} {98}},\ \bibinfo {pages} {063815}
  (\bibinfo {year} {2018})}\BibitemShut {NoStop}%
\bibitem [{\citenamefont {Holstein}\ and\ \citenamefont
  {Primakoff}(1940)}]{Holstein1940}%
  \BibitemOpen
  \bibfield  {author} {\bibinfo {author} {\bibfnamefont {T.}~\bibnamefont
  {Holstein}}\ and\ \bibinfo {author} {\bibfnamefont {H.}~\bibnamefont
  {Primakoff}},\ }\bibfield  {title} {\enquote {\bibinfo {title} {{Field
  Dependence of the Intrinsic Domain Magnetization of a Ferromagnet}},}\ }\href
  {\doibase 10.1103/PhysRev.58.1098} {\bibfield  {journal} {\bibinfo  {journal}
  {Phys. Rev.}\ }\textbf {\bibinfo {volume} {58}},\ \bibinfo {pages}
  {1098--1113} (\bibinfo {year} {1940})}\BibitemShut {NoStop}%
\bibitem [{\citenamefont {Hopfield}(1958)}]{Hopfield1958}%
  \BibitemOpen
  \bibfield  {author} {\bibinfo {author} {\bibfnamefont {J.~J.}\ \bibnamefont
  {Hopfield}},\ }\bibfield  {title} {\enquote {\bibinfo {title} {Theory of the
  contribution of excitons to the complex dielectric constant of crystals},}\
  }\href {\doibase 10.1103/PhysRev.112.1555} {\bibfield  {journal} {\bibinfo
  {journal} {Phys. Rev.}\ }\textbf {\bibinfo {volume} {112}},\ \bibinfo {pages}
  {1555--1567} (\bibinfo {year} {1958})}\BibitemShut {NoStop}%
\bibitem [{\citenamefont {Savasta}\ and\ \citenamefont
  {Girlanda}(1995)}]{Savasta1995}%
  \BibitemOpen
  \bibfield  {author} {\bibinfo {author} {\bibfnamefont {S.}~\bibnamefont
  {Savasta}}\ and\ \bibinfo {author} {\bibfnamefont {R.}~\bibnamefont
  {Girlanda}},\ }\bibfield  {title} {\enquote {\bibinfo {title} {The
  particle-photon interaction in systems descrided by model {H}amiltonians in
  second quantization},}\ }\href
  {https://www.sciencedirect.com/science/article/pii/0038109895002421}
  {\bibfield  {journal} {\bibinfo  {journal} {Solid State Commun.}\ }\textbf
  {\bibinfo {volume} {96}},\ \bibinfo {pages} {517--522} (\bibinfo {year}
  {1995})}\BibitemShut {NoStop}%
\bibitem [{\citenamefont {Sakurai}(1994)}]{Sakurai1994}%
  \BibitemOpen
  \bibfield  {author} {\bibinfo {author} {\bibfnamefont {J.~J.}\ \bibnamefont
  {Sakurai}},\ }\href@noop {} {\emph {\bibinfo {title} {{Modern Quantum
  Mechanics}}}}\ (\bibinfo  {publisher} {Addison-Wesley Publishing Company,
  Inc.},\ \bibinfo {year} {1994})\BibitemShut {NoStop}%
\bibitem [{\citenamefont {Pines}\ and\ \citenamefont
  {Nozieres}(1966)}]{Pines1966}%
  \BibitemOpen
  \bibfield  {author} {\bibinfo {author} {\bibfnamefont {D.}~\bibnamefont
  {Pines}}\ and\ \bibinfo {author} {\bibfnamefont {P}~\bibnamefont
  {Nozieres}},\ }\href@noop {} {\emph {\bibinfo {title} {{The Theory of Quantum
  Liquids}}}}\ (\bibinfo  {publisher} {W. A. Benjamin},\ \bibinfo {address}
  {New York},\ \bibinfo {year} {1966})\BibitemShut {NoStop}%
\bibitem [{\citenamefont {Giuliani}\ and\ \citenamefont
  {Vignale}(2005)}]{Giuliani2005}%
  \BibitemOpen
  \bibfield  {author} {\bibinfo {author} {\bibfnamefont {G.}~\bibnamefont
  {Giuliani}}\ and\ \bibinfo {author} {\bibfnamefont {G.}~\bibnamefont
  {Vignale}},\ }\href@noop {} {\emph {\bibinfo {title} {Quantum theory of the
  electron liquid}}}\ (\bibinfo  {publisher} {Cambridge university press},\
  \bibinfo {year} {2005})\BibitemShut {NoStop}%
\bibitem [{\citenamefont {Andolina}\ \emph {et~al.}(2019)\citenamefont
  {Andolina}, \citenamefont {Pellegrino}, \citenamefont {Giovannetti},
  \citenamefont {MacDonald},\ and\ \citenamefont {Polini}}]{Andolina2019}%
  \BibitemOpen
  \bibfield  {author} {\bibinfo {author} {\bibfnamefont {G.~M.}\ \bibnamefont
  {Andolina}}, \bibinfo {author} {\bibfnamefont {F.~M.~D.}\ \bibnamefont
  {Pellegrino}}, \bibinfo {author} {\bibfnamefont {V.}~\bibnamefont
  {Giovannetti}}, \bibinfo {author} {\bibfnamefont {A.~H.}\ \bibnamefont
  {MacDonald}}, \ and\ \bibinfo {author} {\bibfnamefont {M.}~\bibnamefont
  {Polini}},\ }\bibfield  {title} {\enquote {\bibinfo {title} {Cavity quantum
  electrodynamics of strongly correlated electron systems: A no-go theorem for
  photon condensation},}\ }\href {\doibase 10.1103/PhysRevB.100.121109}
  {\bibfield  {journal} {\bibinfo  {journal} {Phys. Rev. B}\ }\textbf {\bibinfo
  {volume} {100}},\ \bibinfo {pages} {121109} (\bibinfo {year}
  {2019})}\BibitemShut {NoStop}%
\bibitem [{\citenamefont {Savasta}\ \emph {et~al.}(2020)\citenamefont
  {Savasta}, \citenamefont {Di~Stefano},\ and\ \citenamefont
  {Nori}}]{Savasta2020}%
  \BibitemOpen
  \bibfield  {author} {\bibinfo {author} {\bibfnamefont {S.}~\bibnamefont
  {Savasta}}, \bibinfo {author} {\bibfnamefont {O.}~\bibnamefont {Di~Stefano}},
  \ and\ \bibinfo {author} {\bibfnamefont {F.}~\bibnamefont {Nori}},\
  }\bibfield  {title} {\enquote {\bibinfo {title} {{TRK sum rule for
  interacting photons}},}\ }\href@noop {} {\bibfield  {journal} {\bibinfo
  {journal} {arXiv:2002.02139}\ } (\bibinfo {year} {2020})}\BibitemShut
  {NoStop}%
\bibitem [{\citenamefont {Keeling}(2007)}]{Keeling2007}%
  \BibitemOpen
  \bibfield  {author} {\bibinfo {author} {\bibfnamefont {J.}~\bibnamefont
  {Keeling}},\ }\bibfield  {title} {\enquote {\bibinfo {title} {{C}oulomb
  interactions, gauge invariance, and phase transitions of the {D}icke
  model},}\ }\href {\doibase 10.1088/0953-8984/19/29/295213} {\bibfield
  {journal} {\bibinfo  {journal} {J. Phys.: Condens. Matter}\ }\textbf
  {\bibinfo {volume} {19}},\ \bibinfo {pages} {295213} (\bibinfo {year}
  {2007})}\BibitemShut {NoStop}%
\bibitem [{\citenamefont {Mu{\~n}oz}\ \emph {et~al.}(2018)\citenamefont
  {Mu{\~n}oz}, \citenamefont {Nori},\ and\ \citenamefont
  {De~Liberato}}]{Munoz2018}%
  \BibitemOpen
  \bibfield  {author} {\bibinfo {author} {\bibfnamefont {C.~S.}\ \bibnamefont
  {Mu{\~n}oz}}, \bibinfo {author} {\bibfnamefont {F.}~\bibnamefont {Nori}}, \
  and\ \bibinfo {author} {\bibfnamefont {S.}~\bibnamefont {De~Liberato}},\
  }\bibfield  {title} {\enquote {\bibinfo {title} {Resolution of superluminal
  signalling in non-perturbative cavity quantum electrodynamics},}\ }\href
  {https://www.nature.com/articles/s41467-018-04339-w} {\bibfield  {journal}
  {\bibinfo  {journal} {Nat. Commun.}\ }\textbf {\bibinfo {volume} {9}},\
  \bibinfo {pages} {1924} (\bibinfo {year} {2018})}\BibitemShut {NoStop}%
\bibitem [{\citenamefont {Sch{\"a}fer}\ \emph {et~al.}(2019)\citenamefont
  {Sch{\"a}fer}, \citenamefont {Ruggenthaler}, \citenamefont {Rokaj},\ and\
  \citenamefont {Rubio}}]{Schafer2019}%
  \BibitemOpen
  \bibfield  {author} {\bibinfo {author} {\bibfnamefont {C.}~\bibnamefont
  {Sch{\"a}fer}}, \bibinfo {author} {\bibfnamefont {M.}~\bibnamefont
  {Ruggenthaler}}, \bibinfo {author} {\bibfnamefont {V.}~\bibnamefont {Rokaj}},
  \ and\ \bibinfo {author} {\bibfnamefont {A.}~\bibnamefont {Rubio}},\
  }\bibfield  {title} {\enquote {\bibinfo {title} {Relevance of the quadratic
  diamagnetic and self-polarization terms in cavity quantum electrodynamics},}\
  }\href {https://arxiv.org/abs/1911.08427} {\bibfield  {journal} {\bibinfo
  {journal} {arXiv:1911.08427}\ } (\bibinfo {year} {2019})}\BibitemShut
  {NoStop}%
\bibitem [{\citenamefont {Cortese}\ \emph {et~al.}(2017)\citenamefont
  {Cortese}, \citenamefont {Garziano},\ and\ \citenamefont
  {De~Liberato}}]{Cortese2017}%
  \BibitemOpen
  \bibfield  {author} {\bibinfo {author} {\bibfnamefont {E.}~\bibnamefont
  {Cortese}}, \bibinfo {author} {\bibfnamefont {L.}~\bibnamefont {Garziano}}, \
  and\ \bibinfo {author} {\bibfnamefont {S.}~\bibnamefont {De~Liberato}},\
  }\bibfield  {title} {\enquote {\bibinfo {title} {Polariton spectrum of the
  {D}icke-{I}sing model},}\ }\href {\doibase 10.1103/PhysRevA.96.053861}
  {\bibfield  {journal} {\bibinfo  {journal} {Phys. Rev. A}\ }\textbf {\bibinfo
  {volume} {96}},\ \bibinfo {pages} {053861} (\bibinfo {year}
  {2017})}\BibitemShut {NoStop}%
\bibitem [{\citenamefont {Nataf}\ and\ \citenamefont
  {Ciuti}(2010)}]{Nataf2010a}%
  \BibitemOpen
  \bibfield  {author} {\bibinfo {author} {\bibfnamefont {P.}~\bibnamefont
  {Nataf}}\ and\ \bibinfo {author} {\bibfnamefont {C.}~\bibnamefont {Ciuti}},\
  }\bibfield  {title} {\enquote {\bibinfo {title} {No-go theorem for
  superradiant quantum phase transitions in cavity {QED} and counter-example in
  circuit {QED}},}\ }\href {https://www.nature.com/articles/ncomms1069}
  {\bibfield  {journal} {\bibinfo  {journal} {Nat. Commun.}\ }\textbf {\bibinfo
  {volume} {1}},\ \bibinfo {pages} {72} (\bibinfo {year} {2010})}\BibitemShut
  {NoStop}%
\bibitem [{\citenamefont {Savasta}\ and\ \citenamefont
  {Girlanda}(1996)}]{Savasta1996}%
  \BibitemOpen
  \bibfield  {author} {\bibinfo {author} {\bibfnamefont {S.}~\bibnamefont
  {Savasta}}\ and\ \bibinfo {author} {\bibfnamefont {R.}~\bibnamefont
  {Girlanda}},\ }\bibfield  {title} {\enquote {\bibinfo {title} {Quantum
  description of the input and output electromagnetic fields in a polarizable
  confined system},}\ }\href {\doibase 10.1103/PhysRevA.53.2716} {\bibfield
  {journal} {\bibinfo  {journal} {Phys. Rev. A}\ }\textbf {\bibinfo {volume}
  {53}},\ \bibinfo {pages} {2716} (\bibinfo {year} {1996})}\BibitemShut
  {NoStop}%
\bibitem [{\citenamefont {Savona}\ \emph {et~al.}(1994)\citenamefont {Savona},
  \citenamefont {Hradil}, \citenamefont {Quattropani},\ and\ \citenamefont
  {Schwendimann}}]{Savona1994}%
  \BibitemOpen
  \bibfield  {author} {\bibinfo {author} {\bibfnamefont {V.}~\bibnamefont
  {Savona}}, \bibinfo {author} {\bibfnamefont {Z.}~\bibnamefont {Hradil}},
  \bibinfo {author} {\bibfnamefont {A.}~\bibnamefont {Quattropani}}, \ and\
  \bibinfo {author} {\bibfnamefont {P.}~\bibnamefont {Schwendimann}},\
  }\bibfield  {title} {\enquote {\bibinfo {title} {{Quantum theory of
  quantum-well polaritons in semiconductor microcavities}},}\ }\href {\doibase
  10.1103/PhysRevB.49.8774} {\bibfield  {journal} {\bibinfo  {journal} {Phys.
  Rev. B}\ }\textbf {\bibinfo {volume} {49}},\ \bibinfo {pages} {8774--8779}
  (\bibinfo {year} {1994})}\BibitemShut {NoStop}%
\bibitem [{\citenamefont {Gubbin}\ \emph
  {et~al.}(2016{\natexlab{a}})\citenamefont {Gubbin}, \citenamefont {Martini},
  \citenamefont {Politi}, \citenamefont {Maier},\ and\ \citenamefont
  {De~Liberato}}]{Gubbin2016a}%
  \BibitemOpen
  \bibfield  {author} {\bibinfo {author} {\bibfnamefont {C.~R.}\ \bibnamefont
  {Gubbin}}, \bibinfo {author} {\bibfnamefont {F.}~\bibnamefont {Martini}},
  \bibinfo {author} {\bibfnamefont {A.}~\bibnamefont {Politi}}, \bibinfo
  {author} {\bibfnamefont {S.~A.}\ \bibnamefont {Maier}}, \ and\ \bibinfo
  {author} {\bibfnamefont {S.}~\bibnamefont {De~Liberato}},\ }\bibfield
  {title} {\enquote {\bibinfo {title} {{Strong and coherent coupling between
  localized and propagating phonon polaritons}},}\ }\href {\doibase
  10.1103/PhysRevLett.116.246402} {\bibfield  {journal} {\bibinfo  {journal}
  {Phys. Rev. Lett.}\ }\textbf {\bibinfo {volume} {116}},\ \bibinfo {pages}
  {246402} (\bibinfo {year} {2016}{\natexlab{a}})}\BibitemShut {NoStop}%
\bibitem [{\citenamefont {Sentef}\ \emph {et~al.}(2018)\citenamefont {Sentef},
  \citenamefont {Ruggenthaler},\ and\ \citenamefont {Rubio}}]{Sentef2018}%
  \BibitemOpen
  \bibfield  {author} {\bibinfo {author} {\bibfnamefont {M.~A.}\ \bibnamefont
  {Sentef}}, \bibinfo {author} {\bibfnamefont {M.}~\bibnamefont
  {Ruggenthaler}}, \ and\ \bibinfo {author} {\bibfnamefont {A.}~\bibnamefont
  {Rubio}},\ }\bibfield  {title} {\enquote {\bibinfo {title} {Cavity
  quantum-electrodynamical polaritonically enhanced electron-phonon coupling
  and its influence on superconductivity},}\ }\href
  {https://advances.sciencemag.org/content/4/11/eaau6969/tab-pdf} {\bibfield
  {journal} {\bibinfo  {journal} {Sci. Adv.}\ }\textbf {\bibinfo {volume}
  {4}},\ \bibinfo {pages} {eaau6969} (\bibinfo {year} {2018})}\BibitemShut
  {NoStop}%
\bibitem [{\citenamefont {Lamowski}\ \emph {et~al.}(2018)\citenamefont
  {Lamowski}, \citenamefont {Mann}, \citenamefont {Hellbach}, \citenamefont
  {Mariani}, \citenamefont {Weick},\ and\ \citenamefont
  {Pauly}}]{Lamowski2018}%
  \BibitemOpen
  \bibfield  {author} {\bibinfo {author} {\bibfnamefont {S.}~\bibnamefont
  {Lamowski}}, \bibinfo {author} {\bibfnamefont {C.-R.}\ \bibnamefont {Mann}},
  \bibinfo {author} {\bibfnamefont {F.}~\bibnamefont {Hellbach}}, \bibinfo
  {author} {\bibfnamefont {E.}~\bibnamefont {Mariani}}, \bibinfo {author}
  {\bibfnamefont {G.}~\bibnamefont {Weick}}, \ and\ \bibinfo {author}
  {\bibfnamefont {F.}~\bibnamefont {Pauly}},\ }\bibfield  {title} {\enquote
  {\bibinfo {title} {Plasmon polaritons in cubic lattices of spherical metallic
  nanoparticles},}\ }\href {\doibase 10.1103/PhysRevB.97.125409} {\bibfield
  {journal} {\bibinfo  {journal} {Phys. Rev. B}\ }\textbf {\bibinfo {volume}
  {97}},\ \bibinfo {pages} {125409} (\bibinfo {year} {2018})}\BibitemShut
  {NoStop}%
\bibitem [{\citenamefont {Gubbin}\ \emph
  {et~al.}(2016{\natexlab{b}})\citenamefont {Gubbin}, \citenamefont {Maier},\
  and\ \citenamefont {De~Liberato}}]{Gubbin2016}%
  \BibitemOpen
  \bibfield  {author} {\bibinfo {author} {\bibfnamefont {C.~R.}\ \bibnamefont
  {Gubbin}}, \bibinfo {author} {\bibfnamefont {S.~A.}\ \bibnamefont {Maier}}, \
  and\ \bibinfo {author} {\bibfnamefont {S.}~\bibnamefont {De~Liberato}},\
  }\bibfield  {title} {\enquote {\bibinfo {title} {{Real-space Hopfield
  diagonalization of inhomogeneous dispersive media}},}\ }\href {\doibase
  10.1103/PhysRevB.94.205301} {\bibfield  {journal} {\bibinfo  {journal} {Phys.
  Rev. B}\ }\textbf {\bibinfo {volume} {94}},\ \bibinfo {pages} {205301}
  (\bibinfo {year} {2016}{\natexlab{b}})}\BibitemShut {NoStop}%
\end{thebibliography}%

\end{document}